\documentclass[prd,twocolumn,showpacs,superscriptaddress,nofootinbib,floatfix,showkeys,10pt]{revtex4-2}
\usepackage{bm}
\usepackage{times}
\usepackage{braket}
\usepackage{amsfonts,amssymb,stmaryrd,latexsym,amsmath}
\usepackage[usenames,dvipsnames]{color}
\usepackage{slashed}
\usepackage{hyperref}
\usepackage{subfigure}
\usepackage{graphicx}
\usepackage{slashed}
\usepackage{orcidlink}
\usepackage{multirow}
\usepackage{appendix}
\usepackage{dashrule}
\usepackage{float}
\allowdisplaybreaks[1]
\usepackage{verbatim}
\usepackage{blkarray} % 提供块状矩阵注释

\newcommand{\Tr}{\text{Tr}}

\def\Pc{{P_{c\bar{c}}}}
\def\SD{{\Sigma_c^{(*)}D^{(*)}}}
\def\SDbar{{\Sigma_c^{(*)}\bar{D}^{(*)}}}

\definecolor{nicered}{rgb}{0.7,0.1,0.1}
\definecolor{nicegreen}{rgb}{0.1,0.5,0.1}
\definecolor{emph}{rgb}{1,0,0}
\definecolor{doub}{rgb}{0.7,0.2,1.0}%{0.93,0.51,0.93}
\definecolor{navyblue}{RGB}{0, 110, 184}

\hypersetup{colorlinks,citecolor=nicegreen,linkcolor=nicered,urlcolor=navyblue}

 \newcommand{\clabel}[2][]{#2}
 \newcommand{\change}[1]{#1}
\begin{document}

% %%%%%%%%%%%%%%%open the reply mode%%%%%%%%%%%%%%%%%%
% \onecolumngrid
% \input{reply2.tex}
% \newpage
% \setcounter{page}{0}
% %%%%%%%%%%%%%%%open the reply mode%%%%%%%%%%%%%%%%%%

    \title{Towards modeling the short-range interactions of hidden/open charm pentaquark molecular states}
    \author{Ru Xu}\affiliation{School of Physics, Sun Yat-sen University, Guangzhou 510275, China}
\author{Lu Meng\,\orcidlink{0000-0001-9791-7138}}\email{lu.meng@rub.de}
	\affiliation{Institut f\"ur Theoretische Physik II, Ruhr-Universit\"at Bochum,  D-44780 Bochum, Germany }

    \author{Hai-Xiang Zhu\,\orcidlink{0009-0008-7084-7924}}\affiliation{School of Physics, Sun Yat-sen University, Guangzhou 510275, China}
    
\author{Ning Li\,\orcidlink{0000-0003-2987-2809}}\email{lining59@mail.sysu.edu.cn}\affiliation{School of Physics, Sun Yat-sen University, Guangzhou 510275, China}
\author{Wei Chen\,\orcidlink{0000-0002-8044-5493}}\email{chenwei29@mail.sysu.edu.cn}
\affiliation{School of Physics, Sun Yat-sen University, Guangzhou 510275, China}
\affiliation{Southern Center for Nuclear-Science Theory (SCNT), Institute of Modern Physics, 
Chinese Academy of Sciences, Huizhou 516000, Guangdong Province, China}

	\begin{abstract}
The hadronic $\Sigma_c^{(*)}\bar{D}^{(*)}$ and $\Sigma_c^{(*)}{D}^{(*)}$ interactions are revisited, with a focus on their short-range parts, motivated by a tension between the interpretations of $P_{c\bar{c}}(4312)$, $P_{c\bar{c}}(4440)$, and $P_{c\bar{c}}(4457)$ in effective field theory (EFT) frameworks and the one-boson-exchange (OBE) model. While the three states can be interpreted as $\Sigma_c\bar{D}^{(*)}$ molecular states within EFT frameworks, this is not feasible in the single-channel OBE model with consistent cutoff.  In this work, the possibility to reconcile OBE model with EFTs by refitting the  $\rho$-, $\omega$- and $\sigma$-exchange interaction is explored and ruled out. It is pointed out that the problem in OBE arises from the strong short-range spin-dependent one-pion-exchange (OPE) interaction and  the fixed signs of other short-range interactions in OBE model also prevent the cancellation. To address this issue, the short-range subtraction strategies within the OBE model are revisited. Two subtraction schemes are explored: removing the delta-function from all interactions and eliminating it only from the pseudoscalar-meson-exchange component. These schemes favor different spin assignments for $P_{c\bar{c}}(4440)$ and $P_{c\bar{c}}(4457)$. Though solving the problem, there is no clear dynamical picture to support the subtraction schemes. We propose a new quark-exchange mechanism motivated by the Pauli principle. Different from the two subtraction schemes in OBE, the quark-exchange mechanism offers an explanation grounded in microscopic dynamics. It is shown that the spin-dependent quark-exchange interaction cancels those from OPE. The differences in the predictions for the spin, isospin, and open-charm partner states of the experimental $P_{c\bar{c}}$ states offer a way to distinguish between the subtracted OBE model and the OBE model with quark-exchange contributions.

	\end{abstract}
	
	\maketitle
	
	\section{Introduction}~\label{sec:intro}

\begin{figure}
    \centering
    \includegraphics[width=0.32\textwidth]{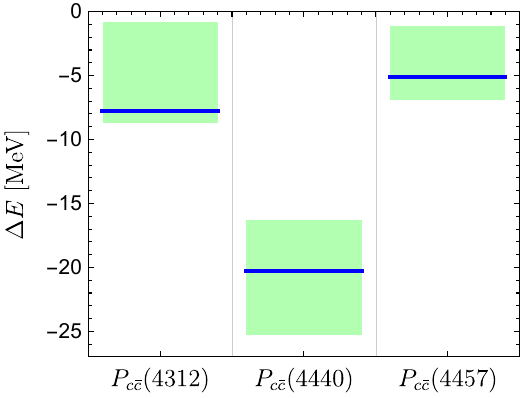}
    \caption{The experimental $\Pc$ masses with respective to two-hadrons thresholds ($\Sigma_c\bar{D}$ threshold for $\Pc(4312)$ and  $\Sigma_c\bar{D}^*$ threshold for $\Pc(4440)$ and $\Pc(4457)$). The isospin-averaged thresholds are used. The blue lines and the green shadows represent the nominal fit and the uncertainties in the LHCb analysis~\cite{LHCb:2019kea}.}
    \label{fig:ex_level}
\end{figure}

 Over the past two decades, a multitude of exotic hadron candidates have been observed in experiments, sparking considerable interest and inspiring numerous theoretical efforts to unravel their underlying nature (see ~\cite{Chen:2016qju,Esposito:2016noz,Guo:2017jvc,Liu:2019zoy,Brambilla:2019esw,Chen:2022asf,Meng:2022ozq,Wang:2025sic} for reviews). Among the various theoretical interpretations proposed, the hadronic molecule picture has garnered significant attention. This framework describes exotic hadrons as loosely bound (or quasi-bound) states of two color-singlet hadrons, analogous to the deuteron in nuclear physics. Identifying hadronic molecules and distinguishing them from other possible configurations, such as compact multiquark states, is a pivotal task in understanding the spectrum of exotic hadrons. In this context, the $T_{cc}(3875)$\cite{LHCb:2021auc, LHCb:2021vvq} and the three $P_{c\bar{c}}$ states ($\Pc(4312)$, $\Pc(4440)$ and $\Pc(4457)$)\cite{LHCb:2015yax, LHCb:2019kea} are particularly promising candidates for hadronic molecules with the following compelling features:
 \begin{itemize}
     \item Manifest multiquark nature: The minimal quark content is $cc\bar{u}\bar{d}$ for $T_{cc}(3875)$ and  $c\bar{c}uud$ for $\Pc$ states, placing them beyond the conventional hadrons;
     \item Proximity to thresholds: Their masses lie close to the thresholds of relevant two-hadron systems, $T_{cc}(3875)$ for $DD^*$ threshold, $\Pc(4312)$ for $\Sigma_c\bar{D}$ threshold and $\Pc(4440)$ and $\Pc(4457)$ for the $\Sigma_c\bar{D}^*$ threshold as shown in Fig~\ref{fig:ex_level}; 
     \item Narrow widths: Their narrow widths indicate that they are likely to be stable particles under the strong interaction once the small coupling to the lower thresholds is neglected;
     \item  Almost stable constituent hadrons: The narrow widths of $\Sigma_c$, $D$ and $D^*$ justify treating these constituents as effectively stable particles in theoretical analyses.
 \end{itemize}

In this work, we focus on the $\Pc$ states. Prior to their experimental observation, theoretical studies had predicted hidden-charm pentaquark states using meson-exchange models~\cite{Wu:2010jy,Yang:2011wz} and the chiral quark model~\cite{Wang:2011rga}. In 2015, the LHCb collaboration reported the first observation of pentaquark states in the $J/\psi p$ invariant mass spectrum from $\Lambda_b^0 \to J/\psi p K^-$ decays~\cite{LHCb:2015yax}. This analysis identified a wide resonance, $\Pc(4380)$, and a narrow resonance, $\Pc(4450)$.  In 2019, an updated analysis incorporating Run II data refined these results~\cite{LHCb:2019kea}. Three narrow resonances were observed with significance above $5\sigma$: $\Pc(4312)$, near the $\Sigma_c \bar{D}$ threshold, and $\Pc(4440)$ and $\Pc(4457)$, near the $\Sigma_c \bar{D}^*$ threshold. The updated analysis did not confirm or refute the existence of $\Pc(4380)$. The relative positions of these states with respect to the thresholds are shown in Fig.~\ref{fig:ex_level}.  These experimental discoveries have spurred extensive theoretical efforts to explore the structure, production mechanisms, and potential partner states of the $\Pc$ resonances~\cite{Chen:2019bip, Chen:2019asm, He:2019ify, Meng:2019ilv, Wang:2019ato, Fernandez-Ramirez:2019koa, He:2019rva,Liu:2019tjn, PavonValderrama:2019nbk, Liu:2019zvb, Du:2019pij, Xiao:2019mvs, Burns:2019iih, Weng:2019ynv, Ali:2019clg, Cao:2019kst, Guo:2019fdo, Lin:2019qiv, Wang:2019hyc, Xiao:2019aya, Xu:2020gjl, Guo:2019kdc, Yamaguchi:2019seo,Yamaguchi:2019vea, Chen:2020pac, Dong:2020nwk, Peng:2020hql, Wang:2021crr, Wang:2021itn, Nakamura:2021qvy, Du:2021fmf, Peng:2022iez, Burns:2022uiv, Yang:2022ezl, Duan:2023dky, Li:2023aui, Zhang:2023czx, Wang:2023eng, Lin:2023ihj}.  The $\Pc$ states were also searched in $J/\psi$ photoproduction ($\gamma p \to J/\psi p$) by the GlueX experiment, which found no evidence for these resonances. This null result and forcoming electron-ion colliders have motivated further theoretical studies of pentaquarks in the context of exclusive $J/\psi$ photoproduction near the threshold~\cite{Winney:2019edt, Xie:2020niw, JointPhysicsAnalysisCenter:2023qgg, Duan:2024hby, Zhang:2024dkm}.  The observations of $\Pc$ states have also inspired predictions of hidden-charm pentaquarks with strangeness~\cite{Chen:2015sxa, Chen:2016ryt, Santopinto:2016pkp,Wang:2019nvm, Shen:2019evi, Xiao:2019gjd} and hidden strange pentaquarks~\cite{Marse-Valera:2023fnv}. Later, the LHCb collaboration reported the evidences of the $P_{c\bar{c}s}(4459)$ and the observation of $P_{c\bar{c}s}(4338)$ in the $J/\psi \Lambda$ invariant mass spectrum from $\Xi_b^- \to J/\psi \Lambda K^-$ decay~\cite{LHCb:2020jpq} and $B^- \to J/\psi \Lambda \bar{p}$ decay~\cite{LHCb:2022ogu}, respectively. These states are near the $\Xi_c \bar{D}^*$ and $\Xi_c \bar{D}$ thresholds, respectively, and have also been the subject of theoretical investigations~\cite{Meng:2022wgl, Chen:2022onm, Feijoo:2022rxf, Giachino:2022pws, Karliner:2022erb, Wang:2022mxy, Yan:2022wuz, Ortega:2022uyu, Nakamura:2022gtu,Wu:2024lud,Chen:2024bre,Li:2025hpd}. Very recently, the Belle and Belle II collaborations reported the evidence of $P_{c\bar{c}s}(4459)$~\cite{Adachi:2025flz}. Beyond hidden-charm pentaquarks, theoretical studies have explored open-charm pentaquarks as well~\cite{Zhu:2019iwm, Chen:2021kad, Shen:2022zvd, Ozdem:2022vip, Wang:2023eng, Lin:2023iww, Liu:2023clr, Wang:2024brl, Duan:2024uuf}. 

In the molecular picture, the $\Pc$ states are (quasi-)bound states of $\Sigma_c \bar{D}$ or $\Sigma_c \bar{D}^*$. Properly establishing the hadronic interactions is critical to this framework. In general, long-range interactions are easy to understand~\cite{Meng:2021jnw,Yang:2025mhg,Meng:2023bmz}, while short-range interactions often remain challenging to determine. Although lattice QCD provides a first-principles approach, there is currently only one lattice study on the $\Pc$ states~\cite{Xing:2022ijm}. Beyond lattice calculations, hadronic interactions are typically determined using two approaches: effective field theories (EFTs) or phenomenological models.

EFT-based interactions are constrained by symmetries, such as chiral symmetry and heavy quark spin symmetry (HQSS), and follow a power-counting scheme based on separated energy scales. In EFT frameworks, including pionless EFT~\cite{Liu:2019tjn} and chiral EFT~\cite{Meng:2019ilv, Wang:2019ato, PavonValderrama:2019nbk, Du:2019pij, Du:2021fmf}, the unknown short-range interactions are parameterized by low-energy constants (LECs), which are determined from experimental data. In the HQSS limit, seven isospin-$\frac{1}{2}$ $\SDbar$ states exist, characterized by different total spin structures: $\Sigma_c \bar{D}(\frac{1}{2})$, $\Sigma_c \bar{D}^*(\frac{1}{2})$, $\Sigma_c \bar{D}^*(\frac{3}{2})$, $\Sigma_c^* \bar{D}(\frac{3}{2})$, $\Sigma_c^* \bar{D}^*(\frac{1}{2})$, $\Sigma_c^* \bar{D}^*(\frac{3}{2})$, and $\Sigma_c^* \bar{D}^*(\frac{5}{2})$, where the values in parentheses represent the total spin $J$. The interaction pattern in this framework is straightforward: two LECs describe the central ($C_c$) and spin-dependent ($C_s$) components of the potential:  
\begin{equation}
     V = C_c + C_s \, \bm{s}_{l1} \cdot \bm{s}_{l2},
\end{equation}
where $\bm{s}_{l1} \cdot \bm{s}_{l2}$ is the spin-spin interaction between the light degree of freedom of two particles. The central interaction provides the same attraction for all states while the spin-dependent term introduces splitting based on spin structures. The three observed $\Pc$ states in Fig.~\ref{fig:ex_level} correspond to $\Sigma_c \bar{D}(\frac{1}{2})$, $\Sigma_c \bar{D}^*(\frac{1}{2})$, and $\Sigma_c \bar{D}^*(\frac{3}{2})$ configurations, with spin-dependent matrix element ratios of $0:-2:1$. This implies that the central interaction is sufficient to bind the constituents, while the spin-dependent term induces binding energy splitting. A natural prediction from this framework is that all other $\Sigma_c^* \bar{D}^{(*)}$ states should also form bound states dominated by the central interaction, with small spin-dependent splitting. Within the EFT framework, since the magnitude and sign of the LECs cannot be determined a priori, an interesting question arises~\cite{PavonValderrama:2019nbk}: how do $\Pc(4440)$ and $\Pc(4457)$ correspond to $\Sigma_c \bar{D}^*(\frac{1}{2})$ and $\Sigma_c \bar{D}^*(\frac{3}{2})$? In this work, we refer to the ``normal order" where the spins of $\Pc(4440)$ and $\Pc(4457)$ are $1/2$ and $3/2$, respectively, and the opposite arrangement is termed the ``reverse order". Determining the sign of $C_s$, which dictates the order, is challenging due to the current quality of the experimental data.

Unlike EFT-based frameworks where short-range interactions are determined by fitting experimental or lattice QCD data~\cite{Wang:2020dko,Meng:2021uhz,Meng:2024kkp}, phenomenological models introduce specific dynamical mechanisms to describe interactions. For example, in the one-boson-exchange (OBE) model, interactions are mediated by exchanging $\pi$,$\eta$, $\rho$, $\omega$, and $\sigma$ mesons, with coupling constants often related to those in other hadronic systems. This allows phenomenological models to predict the existence of molecular pentaquarks even before experimental observations~\cite{Wu:2010jy,Yang:2011wz,Wang:2011rga}. However, phenomenological models can become cumbersome when attempting to precisely describe the $\Pc$ spectrum. For instance, in studies based on OBE~\cite{Chen:2019asm,He:2019rva,He:2019ify}, coupled-channel effects must be incorporated, and the cutoff parameters often need to be adjusted separately for systems with different spin structures. This is because the coupling constants are fixed a priori, requiring adjustments in other aspects of the model to match experimental data. 

In principle, the EFT-based framework and phenomenological models should be consistent. When irrelevant degrees of freedom are integrated out in a phenomenological model, the resulting interactions should align with those described by a successful EFT. However, calculations based on the OBE model in Refs.~\cite{Chen:2019asm,He:2019rva,He:2019ify}, which heavily rely on coupled-channel effects, exhibit tension with results derived from EFT frameworks~\cite{Meng:2019ilv, Wang:2019ato, PavonValderrama:2019nbk, Du:2019pij, Du:2021fmf}. In Ref.~\cite{Zhu:2024hgm}, it was shown that the coupling constants of $D^*D$ and $D^*\bar{D}/D^*\bar{D}$ systems in the OBE model can be improved by fitting experimental data, leading to a picture significantly different from the original one. This highlights the possibility to reconcile OBE models with EFTs by refining the coupling constants, which is one of the aim of this work to explore.

The consistency between the OBE model and EFT was also examined in Ref.~\cite{Liu:2019zvb}, where it was shown that achieving agreement requires the removal of short-range interactions in the OBE model, which correspond to the delta-function terms. It should be noticed that the removed terms have already been regulated in the OBE model and their existence does not violate any fundamental physical principles. Moreover, in the high-precision nuclear force models established using the OBE framework, these terms were not eliminated~\cite{Machleidt:1987hj}. Therefore, we believe this adjustment lacks a clear physical interpretation.  A key advantage of phenomenological models like the OBE framework is their broader applicability to other systems. For example, OBE interactions can be extended to describe open charm pentaquark systems using simple G-parity rules.  However, the adjusted OBE interactions for hidden-charm systems in Ref.~\cite{Liu:2019zvb} may not reliably apply to open-charm systems via G-parity rules, as there is no reason that the subtracted parts also satisfy the G-parities rules. The subtraction in Ref.~\cite{Liu:2019zvb} is motivated by EFT for hidden-charm systems, whereas open-charm systems lie far beyond the valid range of it. Although there has been extensive discussion on the adjustment of the OPE interaction in the \(NN\)  force~\cite{Nogga:2005hy,PavonValderrama:2005gu,Epelbaum:2006pt,Epelbaum:2009sd} from the perspective of renormalization in EFT, these studies do not provide clear guidance to establish the interaction in the \(N\bar{N}\) system. In fact, EFTs are blind to the specific mechanism of the short-range interaction. Therefore, to set up a framework that remains valid for both hidden-charm and open-charm systems, it is necessary not only to match the short-range interactions to EFT in the hidden-charm sector, but also to interpret and implement these adjustments through a physically reasonable mechanism. 

In this work, we first determine the coupling constants in the OBE model by fitting the $\Pc$ experimental spectrum to identify the tension between the OBE model and EFT results. Additionally, we demonstrate that there are multiple ways to modify the short-range interactions to reconcile the OBE model with EFT results. Notably, we find that different adjustments favor different mass orderings for the two $\Sigma_c\bar{D}^*$ states. Furthermore, we propose that the adjustments to the short-range interactions in OBE model may originate from quark exchange interactions, driven by the antisymmetry requirement for identical quarks. While the current data do not allow us to fully determine the effects of quark exchange interactions, we suggest that the spectrum of open-charm systems could help distinguish the short-range interactions arising from meson exchanges and those from quark exchange effects. To this end, we provide predictions based on a pure meson-exchange model, paving the way for future comparisons with quark exchange models when sufficient experimental data become available.

The work is organized as follows. In Sec.~\ref{sec:obe}, the analytical form of the OBE interaction is presented. In Sec.~\ref{sec:fit}, the OBE parameters are determined by fitting the $\Pc$ spectrum, and predictions for their HQSS partners, isospin-spin partners, and open charm partners are provided. In Sec.~\ref{sec:qex}, the quark exchange mechanism of  the short-range interaction is proposed. Finally, a brief conclusion is given in Sec.~\ref{sec:concl}.

\section{One-boson-exchange interaction}~\label{sec:obe}

\begin{figure}
    \centering
    \includegraphics[width=0.48\textwidth]{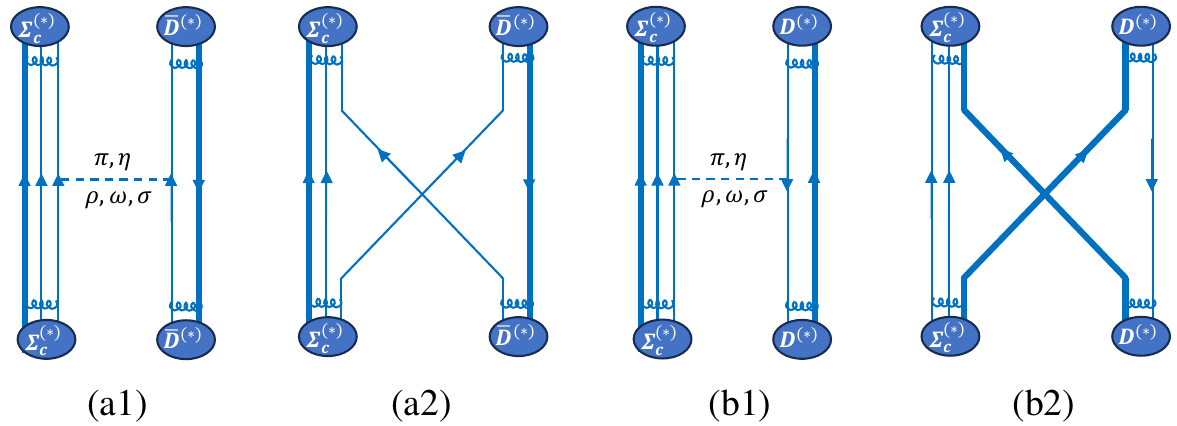}
    \caption{Diagrams illustrating the OBE interaction and quark exchange interaction for the $\SDbar$ and $\SD$ systems. Thick and thin lines represent heavy and light quarks, respectively, with arrows indicating particles and antiparticles.}
    \label{fig:feynman}
\end{figure}

Notably, the OBE model has achieved significant success in elucidating heavy-flavor hadronic molecules~\cite{Liu:2009qhy,Liu:2008fh,Liu:2008xz,Ding:2009vj,Sun:2011uh,Thomas:2008ja,Lee:2009hy,Li:2012cs,Li:2012ss,Dong:2021bvy, Dong:2021juy,Lin:2024qcq}. In this work, the exchanges of $\pi$, $\eta$, $\rho$, $\omega$, and $\sigma$ mesons between $\Sigma_c^{(*)}$ and $\bar{D}^*$ are considered, as illustrated in Fig.~\ref{fig:feynman}(a). The Lagrangians used in this work are shown in Appendix~\ref{app:lag}. The analytical expression for the potential in momentum space can be expressed as  
\begin{equation}    
V = C_{{\text{coupling}}} \times {\cal O}_{s,r} \times \mathcal{O}_{{\text{iso}}},
\end{equation}  
where $C_{\text{coupling}}$, $\mathcal{O}_{s,r}$, and $\mathcal{O}_{\text{iso}}$ represent the coupling constants, spin-spatial components, and isospin operators, respectively. Detailed expressions for these components are provided in Table~\ref{tab:potential}.  Three potential functions in the spin-spatial space are defined as follows:  
\begin{eqnarray}
   && V_{C} = \frac{1}{u^{2}+\bm{q}^{2}}, \quad V_{T} = -\frac{(\bm{q} \cdot \bm{s}_{1})(\bm{q} \cdot \bm{s}_{2})}{u^{2}+\bm{q}^{2}},\\
   && V_{TS} = -\frac{(\bm{q} \cdot \bm{s}_{1})(\bm{q} \cdot \bm{s}_{2})}{u^{2}+\bm{q}^{2}} + \frac{\bm{q}^{2}}{u^{2}+\bm{q}^{2}} \bm{s}_{1} \cdot \bm{s}_{2},\label{eq:VCTTS}
\end{eqnarray}  
where $u$ represents the mass of the exchanged meson. The operator $\bm{s}_i$ depends on the particle type and is defined as:  
\begin{eqnarray}
    \bm{s}_{i} = \begin{cases}
\bm{\sigma} & \text{for } \Sigma_{c},\\
\bm{S}_{rs} & \text{for } \Sigma_{c}^{*},\\
\bm{S}_{v} & \text{for } D^{*},
\end{cases}~\label{eq:spin-operator}
\end{eqnarray}  
where $\bm{S}_{rs}$ and $\bm{S}_v$ are the spin operators for spin-$\frac{3}{2}$ and spin-1 particles, respectively, while $\bm{\sigma}$ represent the Pauli matrices, corresponding to twice the spin operator for spin-$\frac{1}{2}$ particles.  The central interaction arises from vector and scalar meson exchange, while spin-dependent interactions are contributed by vector and pseudoscalar meson exchange. For $S$-wave systems, the replacement $(\bm{q} \cdot \bm{s}_{1})(\bm{q} \cdot \bm{s}_{2}) \to (\bm{s}_1 \cdot \bm{s}_2) \bm{q}^2 / 3$ can be applied.

 \begin{table*}[]
    \centering
        \caption{ Analytical expressions for $\SDbar$ interactions derived from OBE in momentum space. The full potential is the product of $C_{{{coupling}}}$, ${\cal O}_{s,r}$, and $\mathcal{O}_{{{iso}}}$. The coupling constants $\beta$, $g_s$, $g$, and $\lambda$ correspond to the vertices of the $\Sigma_c^{(*)}$ systems, while $\beta_s$, $l_s$, $g_1$, and $\lambda_s$ are associated with the $\bar{D}^{(*)}$ systems. The sign of $C_{{coupling}}$ is determined by quark models. Expressions for $V_C$, $V_T$, and $V_{TS}$ are provided in Eq.~\eqref{eq:VCTTS}. The $\bm{I}_i$ denote isospin operators. The $\SD$ interaction can be derived using the G-parity rule, with the G-parities listed in the second column. Expressions in coordinate space can be easily obtained via Fourier transformation, see Appendix~\ref{app:fourier}. }
    \label{tab:potential}
\begin{tabular*}{\hsize}{@{}@{\extracolsep{\fill}}cccccccc@{}}
\hline 
\hline 
\multirow{2}{*}{Mesons} & \multirow{2}{*}{G-parity} & \multicolumn{2}{c}{Coupling} & Spin-spatial & \multicolumn{3}{c}{Isospin}\tabularnewline
\cline{3-8} \cline{4-8} \cline{5-8} \cline{6-8} \cline{7-8} \cline{8-8} 
 &  & $C_{coupling}$ & sign$(C_{coupling})$  & $\mathcal{O}_{s,r}$ & ${\cal O}_{iso}$ & $I=\frac{1}{2}$ & $I=\frac{3}{2}$\tabularnewline
\hline 
$\rho$ & $+1$ & \multirow{2}{*}{$-\frac{\beta\beta_{s}g_{v}^{2}}{2}$} & \multirow{2}{*}{1} & \multirow{2}{*}{$V_{C}$} & $\bm{I}_{1}\cdot\bm{I}_{2}$ & $-1$ & $\frac{1}{2}$\tabularnewline
$\omega$ & $-1$ &  &  &  & $\frac{1}{2}$ & $\frac{1}{2}$ & $\frac{1}{2}$\tabularnewline
$\sigma$ & $+1$ & $-l_{s}g_{s}$ & -1 & $V_{C}$ & 1 & 1 & 1\tabularnewline
\hline 
$\pi$ & $-1$ & \multirow{2}{*}{$\frac{gg_{1}}{f_{\pi}^{2}}$} & \multirow{2}{*}{1} & \multirow{2}{*}{$V_{T}(\bm{s}_{1},\bm{s}_{2})$} & $\bm{I}_{1}\cdot\bm{I}_{2}$ & $-1$ & $\frac{1}{2}$\tabularnewline
$\eta$ & $+1$ &  &  &  & $\frac{1}{6}$ & $\frac{1}{6}$ & $\frac{1}{6}$\tabularnewline
$\rho$ & $+1$ & \multirow{2}{*}{$-\frac{2}{3}\lambda\lambda_{s}g_{v}^{2}$} & \multirow{2}{*}{-1} & \multirow{2}{*}{$V_{TS}(\bm{s}_{1},\bm{s}_{2})$} & $\bm{I}_{1}\cdot\bm{I}_{2}$ & $-1$ & $\frac{1}{2}$\tabularnewline
$\omega$ & $-1$ &  &  &  & $\frac{1}{2}$ & $\frac{1}{2}$ & $\frac{1}{2}$\tabularnewline
\hline 
\hline 
\end{tabular*}
\end{table*}

In Table~\ref{tab:potential}, the coupling constants $\beta$, $g_s$, $g$, and $\lambda$ correspond to the vertices of the $\bar{D}^{(*)}$ mesons, while $\beta_s$, $l_s$, $g_1$, and $\lambda_s$ are associated with the $\Sigma_c^{(*)}$ baryons. It should be noted that HQSS has been employed to unify the coupling constants of $\Sigma_c$ and $\Sigma_c^*$ as well as $\bar{D}$ and $\bar{D}^*$.  Among these coupling constants, only $g$ for the $\bar{D}^{(*)}\bar{D}^{(*)}\pi$ vertex can be extracted from the decay process $D^* \to D\pi$. For $\Sigma_c^{(*)}\Sigma_c^{(*)}\pi$ coupling $g_1$, there are three main choices in the literature. In Refs.~\cite{Liu:2011xc, Meguro:2011nr, Wang:2018cre, Wang:2018gpl, Meng:2018gan, Meng:2019ilv, Chen:2019asm, Wang:2019nvm, Chen:2021kad, Chen:2024xks}, $g_1$ is taken as $ g_1 = 0.92-0.98 $, where the $\Sigma_c\Sigma_c\pi$ vertices are related to the $\Sigma_c^{(*)} \to \Lambda_c\pi$ process via quark models. In Refs.~\cite{Detmold:2011bp, Detmold:2012ge}, lattice calculations in the static heavy quark limit yield $ g_1 \approx 0.56 $, a value also adopted in Refs.~\cite{Liu:2019zvb, Du:2019pij, Du:2021fmf}. This result is more reasonable for bottom systems than for charm systems due to uncertainty of the static heavy quark limit. The third option is another lattice result based on an axial charge calculation~\cite{Alexandrou:2016xok}, which extracts $ g_1 \approx 0.71 $ for the $\Sigma_c$ system. This value has been used to investigate the scattering of charmed baryons on nucleons~\cite{Haidenbauer:2017dua}. In this work, we adopt the third value derived from first-principles calculations. The remaining coupling constants can only be estimated using phenomenological models~\cite{Chen:2019asm,Liu:2019zvb}. In this work, we will not adopt the values from the literature but instead determine them by fitting the $\Pc$ spectrum. The values from Ref.~\cite{Chen:2019asm}, presented in Table~\ref{tab:lag_para}, will serve as a baseline. To this end, we introduce variable scaling factors for the baseline values of the coupling constants:  
\begin{equation}
    \beta\beta_s \to R_{\beta} \beta\beta_s, \quad l_s g_s \to R_s l_s g_s, \quad \lambda\lambda_s \to R_{\lambda} \lambda\lambda_s. \label{eq:coupling_scale}
\end{equation}  
It is important to note that while the precise values of these coupling constants are challenging to determine \textit{a priori}, the signs of $C_{\text{coupling}}$ can be reliably deduced from quark models, as shown in the fourth column of Table~\ref{tab:lag_para}. Consequently, in the subsequent fitting procedure, we constrain the signs of the coupling constants based on the quark model and only seek solutions for positive values of $R_{\lambda}$, $R_\beta$, and $R_s$.

The potential in coordinate space can be readily derived from its counterpart in momentum space using the following transformations: 
\begin{eqnarray}
\frac{1}{u^{2}+\bm{q}^{2}} &\rightarrow& \frac{e^{-ur}}{4\pi r}, \\
\frac{\bm{q}^{2}}{u^{2}+\bm{q}^{2}} &=& 1 - \frac{u^{2}}{u^{2}+\bm{q}^{2}} \to \delta^{3}(\bm{r}) - \frac{u^{2}e^{-ur}}{4\pi r},~\label{eq:showcasedelta1} \\
\frac{\bm{q}_{i}\bm{q}_{j}}{u^{2}+\bm{q}^{2}} &\rightarrow& -\frac{e^{-ur}}{4\pi r} \left(\frac{u^{2}}{3} + \frac{u}{r} + \frac{1}{r^{2}}\right) T_{ij} \nonumber \\ 
&& + \frac{1}{3} \left(\delta^{3}(\bm{r}) - \frac{u^{2}e^{-ur}}{4\pi r}\right) \delta_{ij}, ~\label{eq:showcasedelta2}
\end{eqnarray}  
with  
\begin{equation}
    T_{ij} = \frac{3r_{i}r_{j}}{r^{2}} - \delta_{ij}, 
\end{equation}  
representing the tensor force term.  Two singularities in the coordinate-space potentials can be readily identified: the delta-function term and the $1/r^3$ term in the tensor force. These singularities arise due to short-range interactions, which lie beyond the valid range of the OBE interaction. A practical approach to mitigate these contributions from the high-momentum region is to apply a regulator:  
\begin{eqnarray}
    V(\bm{q},u) &\to& V(\bm{q},u) F^{2}(u, \Lambda, \bm{q}^{2}), \nonumber \\  
    F(u, \Lambda, \bm{q}^{2}) &=& \frac{\Lambda^{2} - u^{2}}{\Lambda^{2} + \bm{q}^{2}}, \label{eq:regulator}
\end{eqnarray}  
where $ F(u, \Lambda, \bm{q}^{2}) $ is the regulator function. The Fourier transform of the regulated potential is provided in Appendix~\ref{app:fourier}. After regularization, the coordinate-space potential no longer exhibits either the delta-function singularity or the $1/r^3$ singularity. {The regulator in Eq.~\eqref{eq:regulator} can effectively modify the coupling. In principle, independent cutoff parameters could be chosen for different vertices to adjust the interaction. However, since the regulator satisfies \( F(u, \Lambda, \bm{q}^2) < 1 \), it can only soften the interaction and cannot enhance it. To address scenarios where the baseline interaction requires enhancement, we introduce an additional factor in Eq.~\eqref{eq:coupling_scale} and apply a uniform cutoff parameter across all vertices.  }

The $\SD$ interaction within the OBE model can be easily derived from the $\SDbar$ interaction using the G-parity rule. The G-parities of the exchanged mesons are provided in Table~\ref{tab:potential}.

\begin{table}[]
    \centering
        \caption{Hadron masses and baseline coupling constants~\cite{Chen:2019asm,ParticleDataGroup:2024cfk}. All masses and $f_\pi$ are given in units of GeV, while $\lambda$ and $\lambda_s$ are expressed in $\text{GeV}^{-1}$. The present values of $g$ and $f_\pi$ have been updated based on recent experimental results, leading to differences compared to those in Ref.~\cite{Chen:2019asm}. The $g_1$ is take from lattice QCD simulations~\cite{Alexandrou:2016xok}. The values of $g_sl_s$, $\lambda\lambda_s$, and $\beta\beta_s$ are used as baseline parameters, where a rescaling factor can be applied, as specified in Eq.~\eqref{eq:coupling_scale}.  }
    \label{tab:lag_para}    
\begin{tabular*}{\hsize}{@{}@{\extracolsep{\fill}}ccccccc@{}}
\hline 
\hline 
$m_{\pi}$ & $m_{\eta}$ & $m_{\rho}$ & $m_{\omega}$ & $m_{\sigma}$ & &\tabularnewline
0.137 & 0.548 & 0.775 & 0.783 & 0.600 & &\tabularnewline
\hline 
$f_{\pi}$ & $g$ & $\beta$ & $\lambda$ & $g_{s}$ & $m_{D}$ & $m_{D^{*}}$\tabularnewline
0.130 & 0.57 & 0.9 & 0.56 & 0.76 & 1.867 & 2.009 \tabularnewline
\hline 
$g_{v}$ & $g_{1}$ & $\beta_{s}$ & $\lambda_{s}$ & $l_{s}$ & $m_{\Sigma_{c}}$ & $m_{\Sigma_{c}^{*}}$\tabularnewline
5.8 & 0.71 & -1.74 & 3.31 & 6.2 & 2.453 & 2.518\tabularnewline
\hline 
\hline
\end{tabular*}
\end{table}

\section{Fit the $\Pc$ spectrum}~\label{sec:fit}

To determine $ R_\beta $, $ R_\lambda $, and $ R_s $ in Eq.~\eqref{eq:coupling_scale}, we fit the spectrum of the $\Pc(4312)$, $\Pc(4440)$, and $\Pc(4457)$ states. Since the number of parameters matches the number of inputs, this becomes a numerical solution problem. To account for the uncertainties in the $\Pc$ states, we consider three values for each state: the upper limit, the nominal value, and the lower limit. Additionally, we impose the constraint that the binding energy order must remain consistent with the nominal results. This setup leads to a maximum of 27 possible combinations of inputs as follows:  
\begin{eqnarray}
  && \Delta E (\Pc(4312)) \in \{-0.8, -7.8, -8.7\}~\text{MeV}, \nonumber\\
  && \Delta E(\Pc(4440)) \in \{-16.3, -20.3, -25.3\}~\text{MeV},~\label{eq:inputbinding} \\
  && \Delta E(\Pc(4457)) \in \{-1.1, -5.1, -6.9\}~\text{MeV}, \nonumber\\
  && \Delta E(\Pc(4440)) < \Delta E(\Pc(4312)) < \Delta E(\Pc(4457)).\nonumber
\end{eqnarray}

The cutoff parameter in the potential is varied from 1.0 GeV to 1.4 GeV. To align with simple EFT calculations, we do not consider coupled-channel effects but  include only the S-wave interactions.

\subsection{Interaction without subtraction}~\label{sec:Vnosub}

\begin{figure*}
    \centering
    \includegraphics[width=0.85\textwidth]{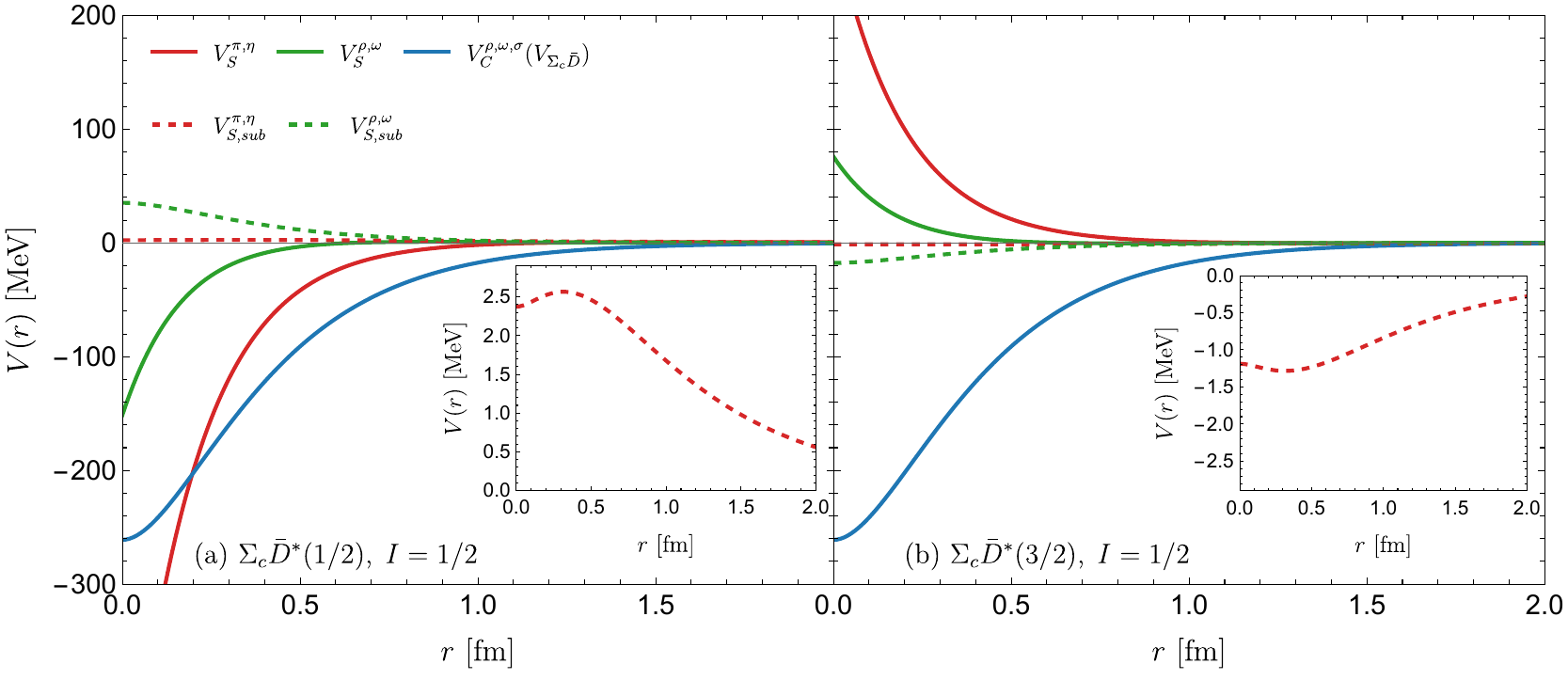}
    \caption{The potentials in the coordinate space. $V_C^{\rho,\omega,\sigma}$ denotes the central potential while $V_S^{\pi,\eta}$ and $V_S^{\rho,\omega}$ denote the spin-dependent interaction from the respective pseudoscalar-meson and vector-meson exchanges. $V_{S,sub}^i$ is the spin-dependent interaction with the delta term subtracted. The central interaction is adjusted to get the experimental $\Pc(4312)$ binding energy.  }
    \label{fig:ptl}
\end{figure*}

\begin{figure}
    \centering
    \includegraphics[width=0.40\textwidth]{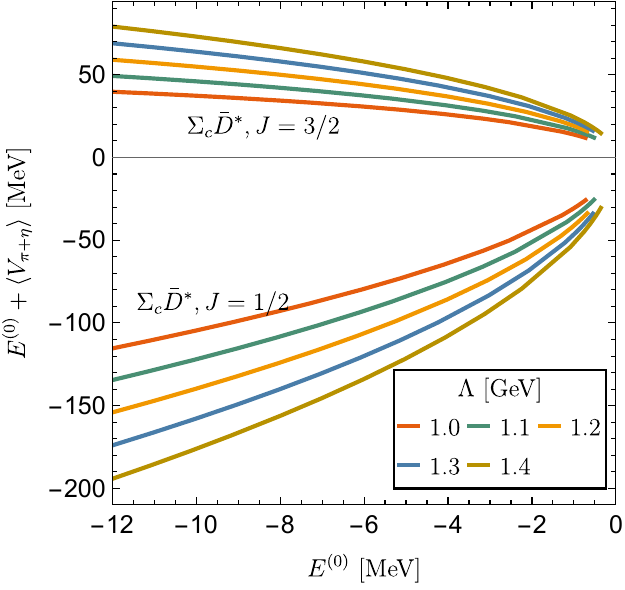}
    \caption{Binding energy splittings of the $\Sigma_c\bar{D}^*$ system within a perturbation theory. $E^{(0)}$ denotes 
    the binding energy of $\Sigma_c\bar{D}$ while $E^{(0)} + \left<V_{\pi +\eta}\right>$ is the binding energy of the 
    $\Sigma_c\bar{D}^*$ system in a perturbation theory.}
    \label{fig:splitting}
\end{figure}
We first determine the parameters using the interaction without subtraction. By varying $R_\beta$, $R_\lambda$, and $R_s$ to search for solutions for any input in Eq.~\eqref{eq:inputbinding}, we find none. During this process, $R_\beta$, $R_\lambda$, and $R_s$ are constrained to be positive to ensure the coupling constants align with the quark model predictions. This result is understandable. As mentioned in Sec.~\ref{sec:intro}, fitting the $\Pc$ spectrum requires a dominant central interaction and a small spin-dependent interaction to account for the binding energy splittings. However, we will see that in the non-subtraction scheme, the one-pion-exchange (OPE) interaction generates a significant spin-dependent contribution, preventing obtaining the solution.

In general, we decompose the S-wave interaction as
\begin{equation}
    V=V_C^{\rho,\omega,\sigma}+V_S^{\rho,\omega}+V_S^{\pi\eta},
\end{equation}
where $V_C$ and $V_S$ are central and spin-dependent interaction, respectively. The superscripts label the exchanged mesons. Focusing first on $\Pc(4312)$, where the spin-dependent interaction is absent, the central interaction arises from $\rho$-, $\omega$-, and $\sigma$-exchange (see Table~\ref{tab:potential}). Since the $\rho$ and $\omega$ masses are nearly identical, the interaction's sign is determined by $\rho$-exchange, which has a larger isospin factor. Consequently, the central interactions from vector meson and scalar meson exchange are both attractive. Their strengths can be adjusted by varying $R_\beta$ or $R_s$ to match the $\Pc(4312)$ binding energy. We present the central potential $V_C^{\rho,\omega,\sigma}$ in Fig.~\ref{fig:ptl} permitting the nominal binding energy of $\Pc(4312)$.

For spin-dependent interactions of $\Sigma_c\bar{D}^*$ systems, the sign of the vector-meson-exchange contribution is also determined by $\rho$-exchange. In the pseudoscalar-meson-exchange interaction, $\eta$-exchange plays a minor role due to its small isospin factor. In some studies, the $\eta$-exchange interaction is neglected, and we find no qualitative changes when doing so. The dominant contribution comes from the one-pion-exchange (OPE) interaction, which primarily governs the spin-dependent pseudoscalar-meson interaction. As shown in Table~\ref{tab:potential}, the spin-dependent OPE and vector-meson-exchange interactions share the same sign. Figure~\ref{fig:ptl} reveals similar results,
\begin{eqnarray}
    \text{sign}(V_S^{\rho,\omega})=\text{sign}(V_S^{\pi,\eta}).
\end{eqnarray}
 The magnitude of the total spin-dependent interaction $V_S$ is thus lower-bounded by the $V_S^{\pi,\eta}$, whose coupling is fixed. Notably, the $V_S^{\pi,\eta}$ is comparable to, or even exceeds, the central interaction $V_C^{\rho,\omega,\sigma}$ in magnitude. This strong spin-dependent interaction would result in a deeply bound $\Sigma_c\bar{D}^*(1/2)$ state and an unbound $\Sigma_c\bar{D}^*(3/2)$ state, both of which contradict the target binding energies in Eq.~\eqref{eq:inputbinding}.

We can prove the above statement via a simple calculation based on perturbation theory. We could set $V_S^{\rho,\omega}=0$ to check the result in the limit. Compared the central interaction, we treat the $V_S^{\pi,\eta}$ as a perturbation.
\begin{eqnarray}
  &  (H_0+V_C)|\psi^{(0)}\rangle =E^{(0)}|\psi^{(0)}\rangle,\nonumber\\
   & E^{(1)}=E^{(0)}+\langle \psi^{(0)}|V_S^{\pi,\eta}|\psi^{(0)}\rangle,
\end{eqnarray}
where the zero-order eigenvalue is just the binding energy of the $\Sigma_c\bar{D}$ system, namely, $\Pc(4312)$. In Fig.~\ref{fig:splitting}, we plot the first-order energy of $\Sigma_c\bar{D}^*$ systems versus the binding energy of $\Sigma_c\bar{D}$ system. One can see clearly, due to the strong spin-dependent interaction, the $\Sigma_c\bar{D}^*(3/2)$ is unbound whereas the $\Sigma_c\bar{D}^*(1/2)$ is a deeply bound state with binding energy mostly larger than $-30$ MeV. Therefore, we conclude that the failure to obtain a solution in the non-subtracted interaction stems from the large spin-dependent interaction in the one-pion exchange. This issue cannot be resolved by adjusting the vector-meson-exchange interaction, as it would only further enhance the spin-dependent interaction. We also test smaller pionic coupling constants, $g_1$, as used in Refs.~\cite{Detmold:2011bp,Detmold:2012ge}, which is insufficient to resolve this problem.

\subsection{Interaction with subtraction}~\label{sec:Vsub}

The above problem that assign $\Pc(4440)$ and $\Pc(4457)$ as $\Sigma_c\bar{D}^*(1/2)$ and $\Sigma_c\bar{D}^*(3/2)$, in any mass order, is hindered by the OPE interaction, were also noted in Refs.~\cite{Burns:2019iih, Liu:2019zvb, Yalikun:2021bfm}. To address this issue, Ref.~\cite{Burns:2019iih} interprets $\Pc(4457)$ as a $\Lambda_c(2595)\bar{D}$ molecular state, while Refs.~\cite{Liu:2019zvb, Yalikun:2021bfm} propose adjusting the short-range interaction. Specifically, the delta-function term is removed entirely in Ref.~\cite{Liu:2019zvb}, and manually reduced by about 80\% in Ref.~\cite{Yalikun:2021bfm}. The effect of the delta-function in OBE were also discussed in Refs.~\cite{Meng:2017fwb, Meng:2017udf,Yalikun:2025ssz}. In Fig.~\ref{fig:ptl}, we show the subtracted potential functions for the pseudoscalar-meson-exchange interaction, $V_{S,\text{sub}}^{\pi,\eta}$, and the spin-dependent vector-meson-exchange interaction, $V_{S,\text{sub}}^{\rho,\omega}$. The strength of $V_{S,\text{sub}}^{\pi,\eta}$ becomes tiny due to the quadratic dependence on the small pion mass in the Yukawa potential, as shown in Eq.~\eqref{eq:showcasedelta1}. Additionally, removing the delta-function term changes the sign of $V_{S,\text{sub}}^{\rho,\omega}$, as indicated in Eq.~\eqref{eq:showcasedelta1}. With the softened OPE interaction, the spin splitting is significantly reduced, resolving the issue.

\begin{figure*}
    \centering
    \includegraphics[width=0.90\textwidth]{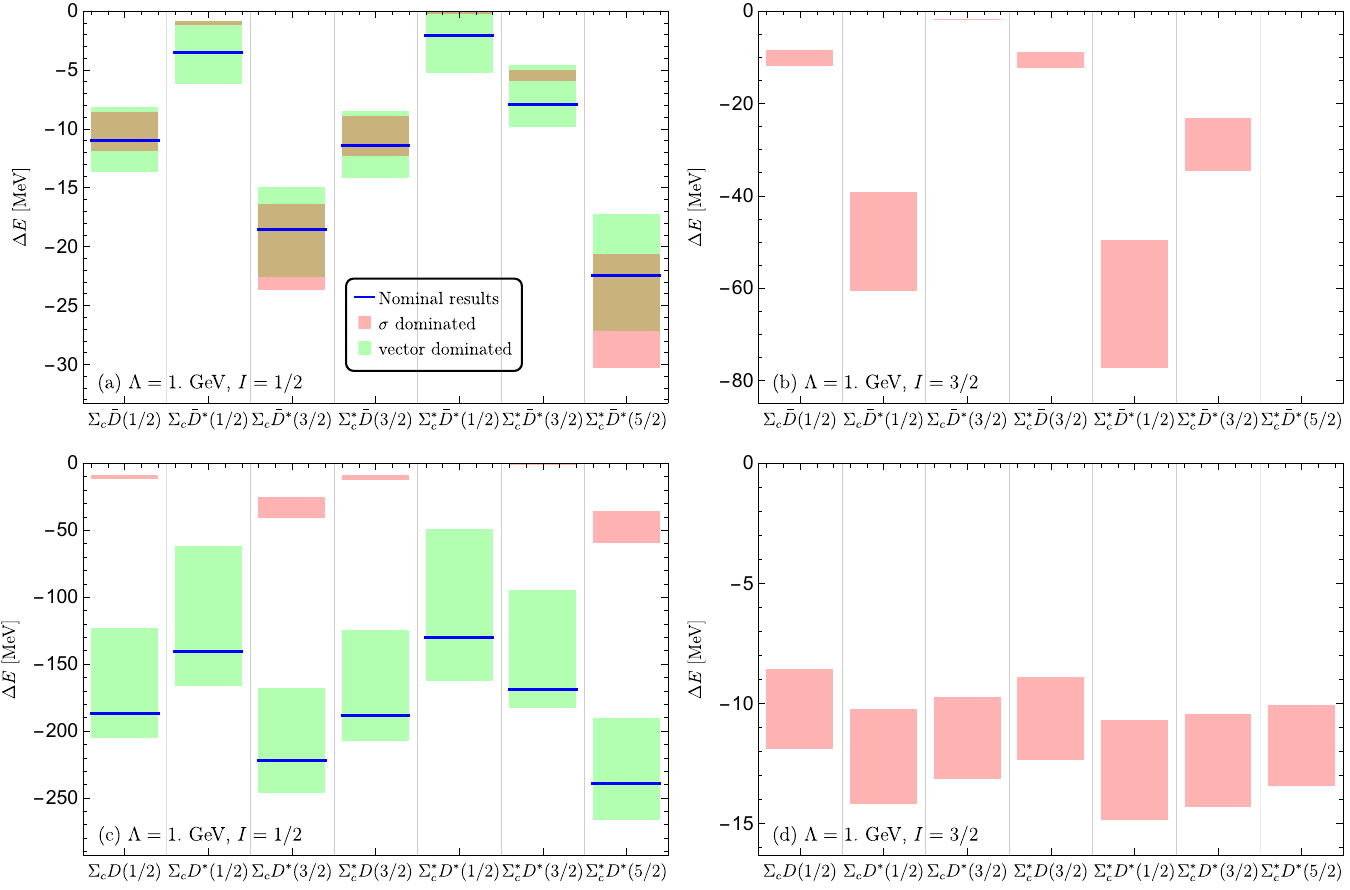}
    \caption{Mass spectrum in the OBE model with $\Lambda=1.0$ GeV for the  subtraction-I scheme. }
    \label{fig:suballreslt}
\end{figure*}

\begin{figure*}
    \centering
    \includegraphics[width=0.90\textwidth]{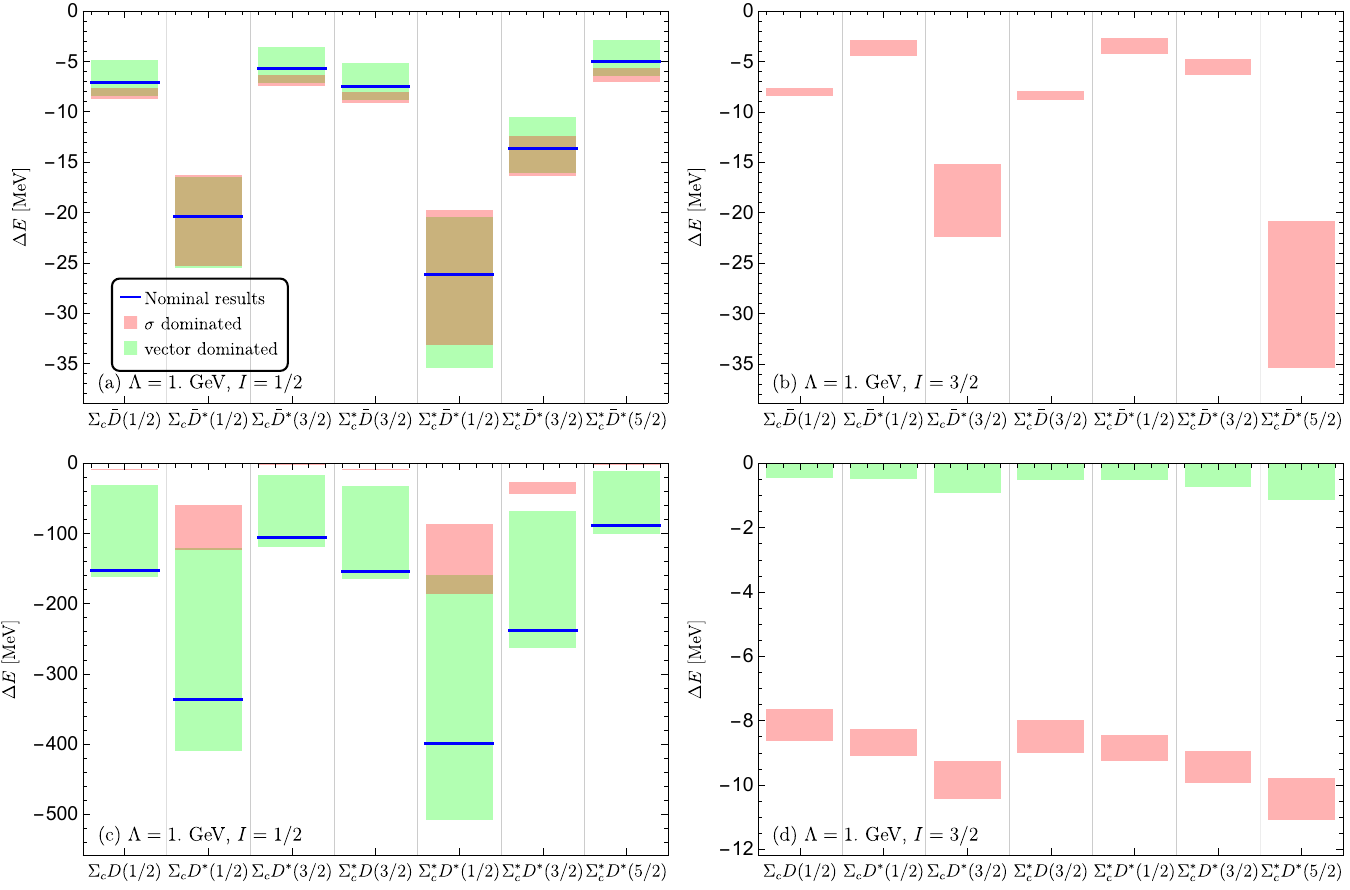}
    \caption{Mass spectrum in the OBE model with $\Lambda=1.0$ GeV for the subtraction-II scheme.}
    \label{fig:subopereslt}
\end{figure*}

In fact, alternative solutions exist if one is not concerned with the specific mechanism of the adjustment. For example, instead of removing the entire delta-function term in the OBE interaction, one could selectively subtract it only from the OPE contribution or the pseudoscalar-meson exchange interaction. In this work, we test two approaches:
\begin{itemize}
    \item subtraction-I: removing the delta-function terms from all OBE interactions;
    \item subtraction-II: selectively removing them only from the pseudoscalar-meson exchange interaction.
    \end{itemize}
We get numerical solutions of $R_\lambda$, $R_\beta$ and $R_s$ within subtraction-I only in the reverse mass order and within subtraction-II only in the normal mass order.  The results using $\Lambda=1.0$ GeV in subtraction-I and subtraction-II are present in Figs.~\ref{fig:suballreslt}  and ~\ref{fig:subopereslt}, respectively. The first case was also present in Ref.~\cite{Liu:2019zvb}, while the later case was exclusively uncovered in that work. More results with different cutoffs $\Lambda=1.2$ and $1.4$ GeV are present in Appendix~\ref{app:cutdepend}. Basically, the results in different cutoffs are consistent qualitatively. All the numerical values of the fitted parameters and the predicted binding energies are available in Zenodo repository~\cite{meng_2025_14958774}.

We present the predictions of the OBE model for the HQSS partner states, isospin partner states, and open-charm partner states of the three experimental $\Pc$ states, as shown in Figs.~\ref{fig:suballreslt} and ~\ref{fig:subopereslt}. In both subtraction schemes, there are four additional bound states within the isospin-1/2 hidden-charm systems. The binding energy of $\Sigma_c^*\bar{D}({3\over 2})$ is approximately $5–15$ MeV, corresponding to the $\Pc(4380)$ state. Notably, the mass order of the three $\Sigma_c^*\bar{D}^*$ states differs between the two subtraction schemes, i.e., decreasing with spin in subtraction-I and increasing with spin in subtraction-II. 

Given the current experimental precision, it is challenging to precisely determine the relative contributions of scalar-exchange and vector-meson-exchange interactions in the central potential. In our analysis, some results favor a vector-meson-dominated central interaction while others support a scalar-meson-dominated one. The predictions for the isospin-3/2 and open-charm systems are particularly sensitive to these specific contributions. To address this, we present the predictions for the isospin partner states and open-charm partner states of the three experimental $\Pc$ states under two schemes: a $\sigma$-dominated $V_C$ and a vector-meson-dominated $V_C$. In the $\sigma$-dominated $V_C$ scheme, all states exhibit bound solutions with binding energies in the range of approximate $5–30$ MeV. This consistency arises because the $\sigma$ interaction is uniform across systems, irrespective of spin, isospin, or the particle/antiparticle nature. In contrast, when the central interaction is dominated by vector-meson exchange, the existence of bound states and their binding energies become very sensitive to spin, isospin, and particle/antiparticle properties. For instance, the isospin-3/2 system lacks bound solutions except for an extremely weakly bound state in the open-charm isospin-3/2 system under subtraction-II. Additionally, the vector-meson-dominated $V_C$ scheme strongly favors deeply bound states for isospin-3/2 hidden-charm systems.

 \section{Quark exchange interaction}~\label{sec:qex}

The general fit based on the OBE model suggests that the short-range interaction needs to be adjusted to align with the three $\Pc$ states. However, the subtraction schemes discussed above are result-driven and not unique. There also exist other adjustment schemes of the short-range interaction other than those in the above section and in Refs.~\cite{Liu:2019zvb, Yalikun:2021bfm}. There is no evidence showing that the adjustment of the short-range interaction could be attributed to the detailed meson-exchanges picture or from other mechanisms. Without a specific picture, it remains uncertain how to relate the subtracted components across systems with different spin, isospin, and particle/antiparticle configurations. In this section, we propose a quark-exchange mechanism to interpret the adjustments required for the short-range interaction, offering a potential explanation grounded in microscopic dynamics. The similar mechanism has been introduced to interpret the repulsion core of the nuclear force~\cite{Oka:1981ri,Oka:1981rj,Fernandez:2019ses} and included in the tetraquark states~\cite{Meng:2023jqk}.

 \begin{figure}
     \centering     
     \includegraphics[width=0.40\textwidth]{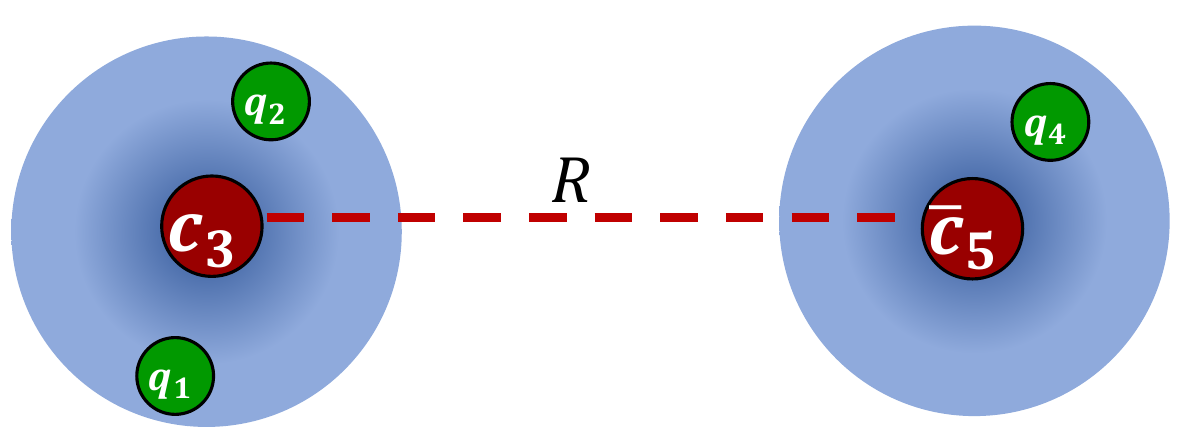}
     \caption{Molecular $\SDbar$ systems in the quark degree of freedom. The $c_1q_2q_3$ and $\bar{c}_4q_5$  form the $\Sigma_c^{(*)}$ and $\bar{D}^{(*)}$, respectively. The quarks are labeled by indices for clarity. }
     \label{fig:qex}
 \end{figure}

As shown in Fig.~\ref{fig:qex}, we consider the static $c$ in $\Sigma_c^{(*)}$ and $\bar{c}$ in $\bar{D}^{(*)}$ with relative distance $R$. The static energy (Born–Oppenheimer potential)  are the good approximation of the hadronic interactions. In the quark level, the wave function of the $\SDbar$ system is expressed as  
\begin{equation}
|\Psi\rangle = |\psi\rangle|\chi\rangle = |\psi(R, 123; 45)\rangle|\chi(123; 45)\rangle,
\end{equation}
where $ |\chi\rangle $ and $ |\psi\rangle $ represent the discrete and spatial wave functions, respectively. The $c_1q_2q_3$ and $\bar{c}_4q_5$  form the $\Sigma_c^{(*)}$ and $\bar{D}^{(*)}$, respectively. At this stage, antisymmetry between the quarks belonging to different hadrons is not imposed, and we label the quarks using indices for clarity.

We can calculate the matrix element of $H$ as follows:
\begin{equation}
H = H_a + H_b + V_{ab},  
\end{equation}
where  
\begin{equation}
\begin{aligned}
    H_a &= T_1 + T_2 + V_{12} + V_{13} + V_{23}, \\  
    H_b &= T_4 + V_{45}, \\  
    V_{ab} &= V_{14} + V_{15} + V_{24} + V_{25} + V_{34} + V_{35}.  
\end{aligned}  
\end{equation}
Here, $T_i$ and $V_{ij}$ represent the kinetic energy and pairwise interaction, respectively. The terms $T_3$ and $T_4$ are neglected in the static $c$ and $\bar{c}$ quark limit. The Hamiltonian is decomposed into three components: $H_a$ and $H_b$, which correspond to the two hadrons, and $V_{ab}$, which accounts for the hadronic interaction between them.  To describe the interaction between two quarks, we introduce the potential  
\begin{equation}
V_{ij} = \epsilon V_{ij}^{OBE} + V_{ij}^{cc} + \epsilon V_{ij}^{cs},  
\end{equation}
where $V_{ij}^{OBE}$, $V_{ij}^{cc}$, and $V_{ij}^{cs}$ represent the OBE, chromoelectric, and chromomagnetic interactions between quarks, respectively. The OBE interaction at the quark level occurs only between light quarks and depends on their spin and isospin. The chromoelectric interaction depends solely on the color charges and is given by  
\begin{equation}
V_{ij}^{cc} = v_{ij}^{cc} O_{ij}^{cc} = v_{ij}^{cc} \left(\frac{\lambda_i}{2} \cdot \frac{\lambda_j}{2}\right),  
\end{equation}
where $O_{ij}^{cc}$ is the chromoelectric operator.  The chromomagnetic interaction is represented by the operator $(\lambda_i \cdot \lambda_j)(\bm{\sigma}_i \cdot \bm{\sigma}_j)$, where it is suppressed by the quark mass. In general, the interaction between quarks is dominated by the chromoelectric interaction. To reflect the subleading contributions of the OBE and chromomagnetic interactions, we introduce the small parameter $\epsilon$.

We can calculate the matrix element of the $H$,
 \begin{eqnarray}
     \langle\Psi|H|\Psi\rangle&&\approx m_{a}+m_{b}+\langle\Psi|V_{ab}|\Psi\rangle \nonumber\\
     &&=m_{a}+m_{b}+\langle\Psi|\epsilon V_{ab}^{OBE}|\Psi\rangle.
 \end{eqnarray}
 In principle, the wave functions of individual hadrons within $ |\Psi\rangle $ should be distorted by the hadronic interaction. However, this effect is expected to be minor due to the molecular approximation. Therefore, we can use the approximation $ H_a|\Psi\rangle \approx m_a|\Psi\rangle $ and $ H_b|\Psi\rangle \approx m_b|\Psi\rangle $, where $ m_a $ and $ m_b $ are the masses of $\Sigma_c^{(*)}$ and $D^{(*)}$, respectively. Furthermore, in the molecular picture, the two clusters remain color singlets, so only the color-independent OBE interaction survives, sandwiched by the molecular wave function. This mechanism is illustrated in Fig.~\ref{fig:feynman} (a1) which aligns with the OBE model at the hadronic level as used in Sec.~\ref{sec:obe}, without considering quark-exchange interactions.

However, the light quarks are identical fermions and must satisfy the Pauli principle, requiring their wave functions to be antisymmetrized. While the two quarks within $\Sigma_c^{(*)}$ already account for exchange antisymmetrization, the light quarks belonging to different hadrons have not been treated in the same way. Incorporating antisymmetrization introduces quark exchange interactions, as illustrated in Fig.~\ref{fig:feynman}(a2). To account for this, we consider a wave function that satisfies exchange antisymmetrization:  
\begin{equation}
|\Psi^{A}\rangle = (1 - P_{14} - P_{24})|\Psi\rangle \equiv |\Psi\rangle + |\Psi^{P}\rangle,    ~\label{eq:anti}  
\end{equation}
where $P_{ij}$ are operators that exchange two quarks. In principle, antisymmetrization would distort the single hadron wave function. However, we treat this distortion as a subleading effect, given that the two clusters are spatially well-separated. Consequently, the antisymmetrized wave function $|\Psi^A\rangle$ is constructed based on original $|\Psi\rangle$.  

Considering antisymmetrization, the matrix elements of the Hamiltonian are expressed as:
 \begin{eqnarray}
\frac{\langle\Psi^{A}|H|\Psi^{A}\rangle}{\langle\Psi^{A}|\Psi^{A}\rangle}&&=\frac{3\langle\Psi^{A}|H|\Psi\rangle}{3\langle\Psi^{A}|\Psi\rangle}=\frac{\langle\Psi^{A}|H|\Psi\rangle}{\langle\Psi^{A}|\Psi\rangle}\nonumber\\
&&=\frac{(m_{a}+m_{b})\langle\Psi^{A}|\Psi\rangle}{\langle\Psi^{A}|\Psi\rangle}+\frac{\langle\Psi^{A}|V_{ab}|\Psi\rangle}{\langle\Psi^{A}|\Psi\rangle}\\
&&=m_a+m_b+{\langle\Psi|\epsilon V_{ab}^{OBE}|\Psi\rangle+\langle\Psi^P| V_{ab}|\Psi\rangle \over \langle\Psi^{A}|\Psi\rangle},\nonumber
 \end{eqnarray}
Here, the first equality relies on the permutation symmetry of the system. The matrix element associated with $V_{ab}$ is divided into two parts, the one without  quark exchange and the other corresponding to quark exchange interaction. For the quark exchange interaction, we primarily consider the dominant chromoelectric interaction:  
\begin{equation}
\langle\Psi^{P}|V_{ij}^{cc}|\Psi\rangle = \sum_{P_{i} \neq 1} (-1)^{P_{i}} \langle P_{i}\chi|O_{ij}^{cc}|\chi\rangle \langle P_{i}\psi|v_{ij}|\psi\rangle,
\end{equation}
where the matrix elements are decomposed into a discrete part and a spatial part. The $ P_i $ denotes the exchange operation, and $ (-1)^{P_i} $ is the corresponding sign factor.  

\begin{table*}
 \centering
\caption{The coefficients related to the quark exchange chromoelectric interactions in Eqs.~\eqref{eq:colormtx} and ~\eqref{eq:quark_exchange}. }~\label{tab:exchagefactor}
  \begin{tabular}{ccccccccc}
\hline 
\hline 
 &  & $\Sigma_{c}\bar{D}(\frac{1}{2})$ & $\Sigma_{c}\bar{D}^{*}(\frac{1}{2})$ & $\Sigma_{c}\bar{D}^{*}(\frac{3}{2})$ & $\Sigma_{c}^{*}\bar{D}(\frac{3}{2})$ & $\Sigma_{c}^{*}\bar{D}^{*}(\frac{1}{2})$ & $\Sigma_{c}^{*}\bar{D}^{*}(\frac{3}{2})$ & $\Sigma_{c}^{*}\bar{D}^{*}(\frac{5}{2})$\tabularnewline
\hline 
$a$ & $I=\frac{1}{2}$ & $-\frac{1}{18}$ & $\frac{1}{54}$ & $-\frac{5}{54}$ & $-\frac{1}{18}$ & $\frac{1}{27}$ & $-\frac{1}{54}$ & $-\frac{1}{9}$\tabularnewline
 & $I=\frac{3}{2}$ & $\frac{1}{9}$ & $-\frac{1}{27}$ & $\frac{5}{27}$ & $\frac{1}{9}$ & $-\frac{2}{27}$ & $\frac{1}{27}$ & $\frac{2}{9}$\tabularnewline
\hline  
$A_{qex}$ & $I=\frac{1}{2}$ & $-\frac{4}{9}$ & $\frac{4}{27}$ & $-\frac{20}{27}$ & $-\frac{4}{9}$ & $\frac{8}{27}$ & $-\frac{4}{27}$ & $-\frac{8}{9}$\tabularnewline
 & $I=\frac{3}{2}$ & $\frac{8}{9}$ & $-\frac{8}{27}$ & $\frac{40}{27}$ & $\frac{8}{9}$ & $-\frac{16}{27}$ & $\frac{8}{27}$ & $\frac{16}{9}$\tabularnewline
\hline 
\hline 
\end{tabular}
 \end{table*}

For the $\SDbar$ systems, the discrete matrix elements read 
\begin{eqnarray}
&\langle P_{i}\chi|O_{ij}^{cc}|\chi\rangle=\nonumber\\
&\begin{blockarray}{rcccccccccc}
  & O_{12}^{cc} & O_{13}^{cc}  & O_{14}^{cc}  & O_{15}^{cc}  & O_{23}^{cc}  & O_{24}^{cc}  & O_{25}^{cc}  & O_{34}^{cc}  & O_{35}^{cc}  & O_{45}^{cc}  \\ % 列注释
\begin{block}{r[cccccccccc]}
1&-\frac{2}{3}&-\frac{2}{3}&0&0&-\frac{2}{3}&0&0&0&0&-\frac{4}{3}\\
-P_{14}&a & a & -2a & 2a & a & a & -a & a & -a & 2a\\
-P_{24}&a & a & a & -a & a & -2a & 2a & a & -a & 2a\\
\end{block}
\end{blockarray},\nonumber\\~\label{eq:colormtx}
\end{eqnarray}
where the rows and columns correspond to interacting operators and exchange operations. The parameter $ a $ depends on the spin and isospin of the system, as detailed in Table~\ref{tab:exchagefactor}. 

The spatial matrix element involves the bra and ket states and potential function. The wave functions have the clustering behavior and the chromoelectric interaction is a short-range interaction, which all have the dominated regions. For the matrix elements involving quark exchange, the bra and ket wave functions always mismatch. The size of the spatial matrix element is determined by how much the dominated regions of potential function match to the bra and ket states. Thus, the spatial matrix elements can be roughly grouped into two categories with hierarchy between them. An example of the larger value category is  
\begin{equation}
\langle P_{14}\psi|v_{24}|\psi\rangle = \langle\psi(R,423;15)|v_{24}|\psi(R,123;45)\rangle \equiv \delta_v,~\label{eq:MM1}    
\end{equation}
 where the dominated region of the potential matches to the clustering behavior of the bra state but mismatches to the ket state. Therefore, we use the $\delta_v$ to label the spatial matrix element with interaction that mismatches to only one state. The small difference among matrix elements in the same category is neglected.  There is another possibility that the dominated region of the interaction mismatches to both states, for example,
\begin{eqnarray}
    \langle P_{14}\psi|v_{14}|\psi\rangle&=&\langle\psi(R,423;15)|v_{14}|\psi(R,123;45)\rangle\equiv\mathcal{O}(\delta_{v}^{2}).\nonumber\\~\label{eq:ex2}
\end{eqnarray}
Thus, we expected the matrix element is further suppressed compared to the first category and is neglected in present estimation.{It should be noted that even when quarks transition from one hadron to another, the interaction must still be regarded as an interaction between two hadrons. This is because the antisymmetrization of quarks, even those from different hadrons, must be taken into account, as demonstrated in Eq.~\eqref{eq:anti}. The wave function of the di-hadron system, denoted as \( |\Psi^A\rangle \), comprises three terms. The expressions in Eqs.~\eqref{eq:MM1} and ~\eqref{eq:ex2} represent matrix elements of different terms, which contribute to the overall term \( \langle\Psi^A|V_{ab}|\Psi^A\rangle \). Therefore, these should be considered as the interaction between two hadrons.}

Therefore, the exchanged matrix element is,
\begin{equation}
    \langle\Psi^{P}|V_{ab}|\Psi\rangle=A_{qex}\delta_{v},~\label{eq:quark_exchange}
\end{equation}
The $A_{eqx}$ is calculated from the discrete matrix elements, which are given in Table~\ref{tab:exchagefactor}. We can summarize the operator form of $A_{eqx}$ as follows:
\begin{equation}
    A_{qex}(s_{1},s_{2},J,I)=\frac{16}{27}(\bm{s}_{1}\cdot\bm{s}_{2}+\frac{3}{2})(\bm{I}_{1}\cdot\bm{I}_{2}+\frac{1}{2}),~\label{eq:quark_exchange_op}
\end{equation}
where $\bm{s_i}$ are defined in Eq.~\eqref{eq:spin-operator}. The final Hamiltonian matrix elements can be got
\begin{eqnarray}
\frac{\langle\Psi^{A}|H|\Psi^{A}\rangle}{\langle\Psi^{A}|\Psi^{A}\rangle}&&\approx m_{a}+m_{b}+\frac{\langle\Psi|\epsilon V_{ab}^{OBE}|\Psi\rangle+A_{qex}\delta_{v}}{1+\delta_{\psi}}\nonumber\\
&&\approx m_{a}+m_{b}+\langle\Psi|\epsilon V_{ab}^{OBE}|\Psi\rangle+A_{qex}\delta_{v},\nonumber\\
\end{eqnarray}
where $\langle\Psi^{A}|\Psi\rangle=1+\delta_{\psi}$ with $\delta_\psi$ being regarded as a small quantity due to the mismatch of the spatial bra and ket wave functions. Compared to the framework without the quark exchange, the quark exchange effect would give rise to an additional interaction term as shown in Eq.~\eqref{eq:quark_exchange}, from the chromoelectric interaction. Since we only focus on the ground state molecules composed of ground state hadrons, we expect there is no node point in the spatial wave function. And the spatial part of chromoelectric interaction, $v_{ij}\sim 1/r$ is also larger than zero. Therefore, one can expect $\delta_{v}>0$. For the isospin-${1\over 2}$, the factor $A_{eqx}$ is
\begin{equation}
    A_{qex}(s_{1},s_{2},J,\frac{1}{2})=-\frac{8}{27}\bm{s}_{1}\cdot\bm{s}_{2}-\frac{4}{9}\,.
\end{equation}
We can see the spin-spin part in this quark-exchange interaction has different signs from those of the OPE interaction. Therefore, with the quark-exchange interaction, the problem in Sec.~\ref{sec:Vnosub} could be solved. The subtraction of the short-range interaction in Sec.~\ref{sec:Vsub} could also stem from the quark-exchange interaction rather than the OBE model.

More intriguingly, as shown in Eq.~\eqref{eq:quark_exchange_op}, the quark-exchange interaction term exhibits a specific spin-isospin structure that cannot be attributed to any single meson-exchange interaction. {For specific systems (e.g., the \( I = \frac{1}{2} \) hidden-charm system), the subtraction within the OBE model can have a similar effect to incorporating the quark-exchange mechanism. However, when extended to general spin, isospin, and open-charm partner states, the subtraction scheme and the quark-exchange model yield distinct differences. }For hidden charm systems, predictions from models combining OBE and quark-exchange interactions are expected to differ from those based solely on OBE. Similarly, for open charm systems, the quark-exchange mechanism could also play a role, as illustrated in Fig.~\ref{fig:feynman} (b2). However, since the exchanged quarks are heavy, the quark-exchange effect may be suppressed by their mass, leading to further deviations from pure OBE predictions. Currently, it is challenging to determine the additional $\delta_v$ parameter beyond those in OBE. The predictions in Figs.~\ref{fig:suballreslt} and~\ref{fig:subopereslt} provide a useful preparation for future comparisons.

\begin{figure}
    \centering
    \includegraphics[width=0.5\textwidth]{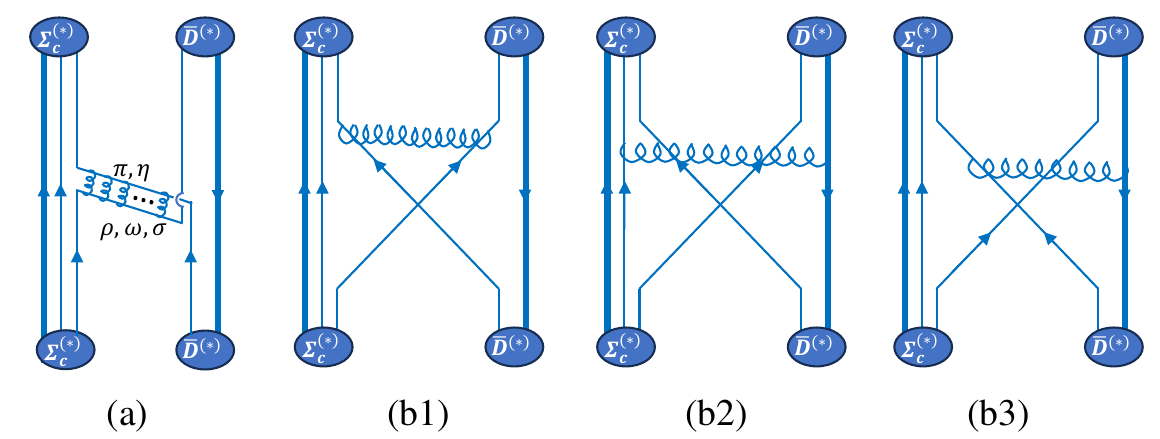}
    \caption{\change{The meson-exchange mechanism at the quark level (a) and the different types of quark exchange interactions (b1)–(b3): (b1) one-gluon exchange between two exchanged quarks, (b2) one-gluon exchange between two unexchanged quarks, and (b3) one-gluon exchange between an exchanged quark and an unexchanged quark. As shown in Eqs.~\eqref{eq:MM1} and~\eqref{eq:ex2}, the contributions from (b1) and (b2) are suppressed compared to (b3) and are therefore neglected. }}
    \label{fig:quarkexchange2}
\end{figure}

\clabel[doublecount]{One might wonder whether the quark-exchange interaction involving one-gluon exchange overlaps with the meson-exchange mechanism, given that the exchanged meson could be interpreted as a quark-antiquark state. As shown in Fig.~\ref{fig:quarkexchange2}(a), meson-exchange diagrams can also facilitate quark interchange between two hadrons. To equate the quark-antiquark interaction with a meson exchange, the interactions between them must be iterated to form a mesonic state.  The quark-exchange diagrams involving one-gluon exchange are presented in Fig.~\ref{fig:quarkexchange2}(b1)–(b3), illustrating different types of interactions. At first glance, the contribution of (b1), involving one-gluon exchange between two exchanged quarks, might appear to be included in the one-meson exchange interaction depicted in subfigure (a). However, in this work, we argue that the contributions from (b1) and (b2) are suppressed compared to (b3), as shown in Eqs.~\eqref{eq:MM1} and~\eqref{eq:ex2}. Only the interaction involving an exchanged quark and an unexchanged quark (b3) are kept. It is evident that subfigure (b3) does not overlap with the one-meson exchange interaction depicted in subfigure (a).  }

\section{conclusion}~\label{sec:concl}
In this work, we investigate the $\SDbar$ and $\SD$ interaction, focusing on its short-range part, motivated by the tension of the OBE model and the EFT frameworks. We first determined the coupling constants in the OBE model by fitting them to the experimental $\Pc$ spectrum, concluding that  it is impossible to reconcile OBE models with EFTs by refitting the  $\rho$-, $\omega$- and $\sigma$-exchange interactions. It is pointed out by a perturbation theory that the problem in OBE arises from the strong short-range spin-dependent one-pion-exchange (OPE) interaction and  the fixed signs of other short-range interactions in OBE model also prevent the cancellation. Once the central interaction is adjusted to permit $\Sigma_c\bar{D}$ to be bound to form $\Pc(4312)$, due to the strong spin-dependent interaction, the $\Sigma_c\bar{D}^*(3/2)$ is unbound whereas the $\Sigma_c\bar{D}^*(1/2)$ is a very deep bound state with binding energy mostly larger than $-30$ MeV.

 To address this issue, the short-range subtraction strategies within the OBE model are revisited, and a new quark-exchange mechanism is proposed. Two subtraction schemes are explored: removing the delta-function from all interactions in OBE and eliminating it only from the pseudoscalar-meson-exchange component. The former one favors the $P_{c\bar{c}}(4440)$ and $P_{c\bar{c}}(4457)$ are respective spin-$3/2$ and spin-$1/2$ bound states while the latter one favors the opposite spin assignment. Though solving the problem, there is no clear dynamical picture to support the subtraction schemes. In contrast, the quark-exchange model motivated by the Pauli principle  offers an explanation grounded in microscopic dynamics, which is also used to interpret the short-range nuclear force.  With certain approximations, we show that the spin-dependent quark-exchange interaction, driven by chromoelectric effects, cancels some of the short-range contributions of the one-pion exchange. This interaction exhibits a distinctive spin-isospin structure unattainable by any single meson-exchange interaction. Furthermore, the quark-exchange mechanisms differ between the $\SDbar$ system, dominated by light-quark exchange, and the $\SD$ system, influenced by heavy-quark exchange. The differences in the predictions for the spin, isospin, and open-charm partner states of the experimental $P_{c\bar{c}}$ states offer a way to distinguish between the subtracted OBE model and the OBE model with quark-exchange contributions.

	\section*{ACKNOWLEDGMENTS}
	
	We thank Zi-Yang Lin and Bo Wang for helpful discussions. This project was supported by the National Natural Science Foundation of China (Grant No. 12175318), ERC NuclearTheory (Grant No. 885150), and the Natural Science Foundation of Guangdong Province of China (Grant No. 2023A15150117). 

    \section*{Data Availability Statement}

The data supporting the findings of this study is available on Zenodo repository~\cite{meng_2025_14958774} or can be obtained directly from the authors upon reasonable request.

\appendix

\section{The Lagrangians of OBE interactions}~\label{app:lag}

Using the HQSS, the pseudoscalar  meson $D$ and vector meson $D^*$ can be formulated in a superfield $\mathcal{H}$~\cite{Georgi:1990cx,Mannel:1990vg,Falk:1991nq,Liu:2008fh,Liu:2009qhy,Li:2012cs,Li:2012ss},
\begin{equation}
        {\cal H}=\frac{1+\slashed{v}}{2}(P_{\mu}^*\gamma^{\mu}-P\gamma_{5}),
\end{equation}
where $P=(D^0,D^+)$ and $P_\mu^*=(D^{*0},D^{*+})_\mu$. The velocity of the heavy meson is denoted by $v=(1,0,0,0)$. Similarly, their antiparticles can be described by the superfield $\tilde{{\cal H}}$, defined as,
\begin{eqnarray}
  \tilde{{\cal H}}=(\tilde{P}_{\mu}^*\gamma^{\mu}-\tilde{P}\gamma_{5})\frac{1-\slashed{v}}{2},
\end{eqnarray}
with $\tilde{P}=(\bar{D}^0,D^-)^\text{T}$ and $\tilde{P}_\mu^*=(\bar{D}^{*0},D^{*-})_\mu^\text{T}$. Here, we choose the convention of charge conjugation, $D\xrightarrow{\mathcal{C}}\bar{D}$ and $D^*\xrightarrow{\mathcal{C}}-\bar{D}^*$, namely $\mathcal{H}\xrightarrow{\mathcal{C}}C^{-1}\tilde{\mathcal{H}}^\text{T}C$, where $C=i\gamma^2\gamma^0$ is the charged conjugation matrix. 
 The conjugations of $\cal{H}$ and $\tilde{\mathcal{H}}$ are defined as $\bar{\mathcal{H}}=\gamma_0\mathcal{H}^\dagger\gamma_0$ and $\bar{\tilde{\mathcal{H}}}=\gamma_0\tilde{\mathcal{H}}^\dagger\gamma_0$. In the HQSS, the $\Sigma_c^*$ and $\Sigma_c$ can be accommodated in a superfield, 
 \begin{equation}
     S_\mu=B_{6\mu}^*-\sqrt{\frac{1}{3}}(\gamma_\mu+v_\mu)\gamma_5B_6,
 \end{equation}
 with the baryon matrix $B_6^{(*)}$
\begin{eqnarray}
     B_6^{(*)}=\begin{bmatrix}\Sigma_c^{(*)++}
 &\frac{\Sigma_c^{(*)+}}{\sqrt{2}}\\
\frac{\Sigma_c^{(*)+}}{\sqrt{2}} &\Sigma_c^{(*)0}
\end{bmatrix}. 
\end{eqnarray}
The conjuration of the superfield is $\bar{S}_\mu=S_\mu^\dagger\gamma_0$. 

The Lagrangians in the OBE model read,

\begin{eqnarray}
    &\mathcal{L}&=g_s\Tr\left[\mathcal{H}\sigma\bar{\mathcal{H}}\right]+ig_a\Tr\left[\mathcal{H}\gamma_\mu\gamma_5\mathcal{A}^\mu\bar{\mathcal{H}}\right]\nonumber\\
    &&+i\beta\Tr\left[\mathcal{H} v_\mu (\mathcal{V}^\mu-\rho^\mu)\bar{\mathcal{H}}\right]+i\lambda\Tr\left[\mathcal{H}\sigma_{\mu\nu}F^{\mu\nu}\bar{\mathcal{H}}\right]\nonumber\\
    &&+g_s\Tr\left[\bar{\tilde{\mathcal{H}}}\sigma\tilde{\mathcal{H}}\right]+ig_a\Tr\left[\bar{\tilde{\mathcal{H}}}\gamma_\mu\gamma_5\mathcal{A}^\mu\tilde{\mathcal{H}}\right]\nonumber\\
    &&-i\beta\Tr\left[\bar{\tilde{\mathcal{H}}} v_\mu (\mathcal{V}^\mu-\rho^\mu)\tilde{\mathcal{H}}\right]+i\lambda\Tr\left[\bar{\tilde{\mathcal{H}}}\sigma_{\mu\nu}F^{\mu\nu}\tilde{\mathcal{H}}\right]\nonumber\\
&&+l_s\Tr\left[\bar{S}_\mu\sigma S^\mu\right]+\frac{3}{2}g_1\varepsilon^{\mu\nu\lambda\kappa}v_\kappa\Tr\left[\bar{S}_\mu \mathcal{A}_\nu S_\lambda\right]\nonumber\\
    &&+i\beta_s\Tr\left[\bar{S}_\mu v_\alpha(\mathcal{V}^\alpha-\rho^\alpha) S^\mu\right]+\lambda_s\Tr\left[\bar{S}_\mu F^{\mu\nu} S_\nu\right],
\end{eqnarray}
where $F^{\mu\nu}=\partial^\mu\rho^\nu-\partial^\nu\rho^\mu-[\rho^\mu,\rho^\nu]$ represents the field strength tensor of vector mesons,  
$\mathcal{V}^\mu$ and $\mathcal{A}^\mu$ are the vector and axial building blocks of pseudoscalar mesons respectively,
\begin{eqnarray}
    &&\mathcal{V}^\mu=\frac{1}{2}[\xi^\dagger,\partial_\mu\xi],\; \mathcal{A}^\mu=\frac{1}{2}\{\xi^\dagger,\partial_\mu\xi\},\;
        % \nonumber\\&
    \xi=\exp(i\mathbb{P}/f_\pi).
\end{eqnarray}
 The pseudoscalar meson matrix $\mathbb{P}$ is defined as
\begin{eqnarray}
\mathbb{P}=
\begin{bmatrix}\frac{\pi^{0}}{\sqrt{2}}+\frac{\eta}{\sqrt{6}} & \pi^{+}\\
\pi^{-} & -\frac{\pi^{0}}{\sqrt{2}}+\frac{\eta}{\sqrt{6}}
\end{bmatrix}.\nonumber\\
\end{eqnarray}
The multiplet of the vector meson fields $\rho^\mu$ is
\begin{eqnarray}
     \rho^\mu=\frac{ig_V}{\sqrt{2}}\begin{bmatrix}\frac{\rho^{0}+\omega}{\sqrt{2}}
 & \rho^{+}\\
\rho^{-} & \frac{-\rho^{0}+\omega}{\sqrt{2}}
\end{bmatrix}^\mu.
\end{eqnarray}

In principle, the iso-triplet $\rho$ ($\pi$) and the iso-singlet $\omega$ ($\eta$) belong to different multiplets under the flavor SU(2) symmetry. However, to reduce the number of coupling constants, we group them into the same matrix, taking into account their relationship within the flavor SU(3) symmetry.

	\section{Fourier transformation}\label{app:fourier}
   
    The potential in the coordinate space is derived from the following Fourier transformation,
    \begin{equation}
        {V}(\bm{r})=\int\frac{d^{3}\bm{q}}{(2\pi)^{3}}e^{i\bm{q}\cdot\bm{r}}{V}(\bm{q}).\label{eq:FTq}
    \end{equation}  
For the specific potential in this work, the Fourier transformation reads: 
\begin{eqnarray}
    \frac{1}{u^{2}+\bm{q}^{2}}F(u,\Lambda,q^{2})^{2}&\to&H_{0}(u,\Lambda,r),\nonumber\\
    \frac{\bm{q}^{2}}{u^{2}+\bm{q}^{2}}F(u,\Lambda,q^{2})^{2}&\to&-H_{1}(u,\Lambda,r),\\
    \frac{q_{i}q_{j}}{u^{2}+\bm{q}^{2}}F(u,\Lambda,q^{2})^{2}&\to&-[H_{3}(u,\Lambda,r)T_{ij}+H_{1}(u,\Lambda,r)\frac{\delta_{ij}}{3}],\nonumber
\end{eqnarray}
where $T_{ij} = \frac{3r_{i}r_{j}}{r^{2}} - \delta_{ij}$. One can also easily obtain the transformation for the case where $V(\bm{p’},\bm{p}) = \tilde{V}(\bm{k})$. The explicit expressions of $H_0$, $H_1$, and $H_3$ are: 
\begin{eqnarray}
  H_{0}(u,\Lambda,r)&=&\frac{u}{4\pi}\left[\frac{e^{-ur}-e^{-\Lambda r}}{ur}-\frac{\Lambda^{2}-u^{2}}{2u\Lambda}e^{-\Lambda r}\right],\nonumber\\
  H_{1}(u,\Lambda,r)&=&\nabla^{2}H_{0}(u,\Lambda,r) \nonumber\\
  &=&\frac{u^{3}}{4\pi}\left[\frac{e^{-ur}-e^{-\Lambda r}}{ur}-\frac{(\Lambda^{2}-u^{2})\Lambda^{2}}{2u^{3}\Lambda}e^{-\Lambda r}\right],\nonumber\\
  H_{3}(u,\Lambda,r)&=&\frac{1}{3}r\frac{\partial}{\partial r}\frac{1}{r}\frac{1}{\partial r}H_{0}(u,\Lambda,r) \nonumber\\
  &=&\frac{u^{3}}{12\pi}\left[ -\frac{e^{-\Lambda r}\Lambda^{2}\left(\frac{3}{\Lambda^{2}r^{2}}+\frac{3}{\Lambda r}+1\right)}{ru^{3}}\right.\nonumber\\&&
  ~~~~~~~~~~~-\frac{e^{-\Lambda r}(\Lambda r+1)\left(\Lambda^{2}-u^{2}\right)}{2ru^{3}}\nonumber\\&&
  ~~~~~~~~~~~\left.+\frac{e^{-ur}\left(\frac{3}{r^{2}u^{2}}+\frac{3}{ru}+1\right)}{ru}\right].  
\end{eqnarray}
Apparently, the regulator smears the delta-function singularity in Eqs.~\eqref{eq:showcasedelta1} and~\eqref{eq:showcasedelta2}. The $1/r^3$ singularity in the tensor force in Eq.~\eqref{eq:showcasedelta2} is also regulated, rendering for $r\to 0$
\begin{equation}
    H_{3}(u,\Lambda,r)\to\frac{\Lambda^{4}+u^{4}-2\Lambda^{2}u^{2}}{96\pi}r+\mathcal{O}(r^{2}).
\end{equation}

\section{Cutoff dependence}~\label{app:cutdepend}
\begin{figure*}[h]
    \centering
    \includegraphics[width=0.90\textwidth]{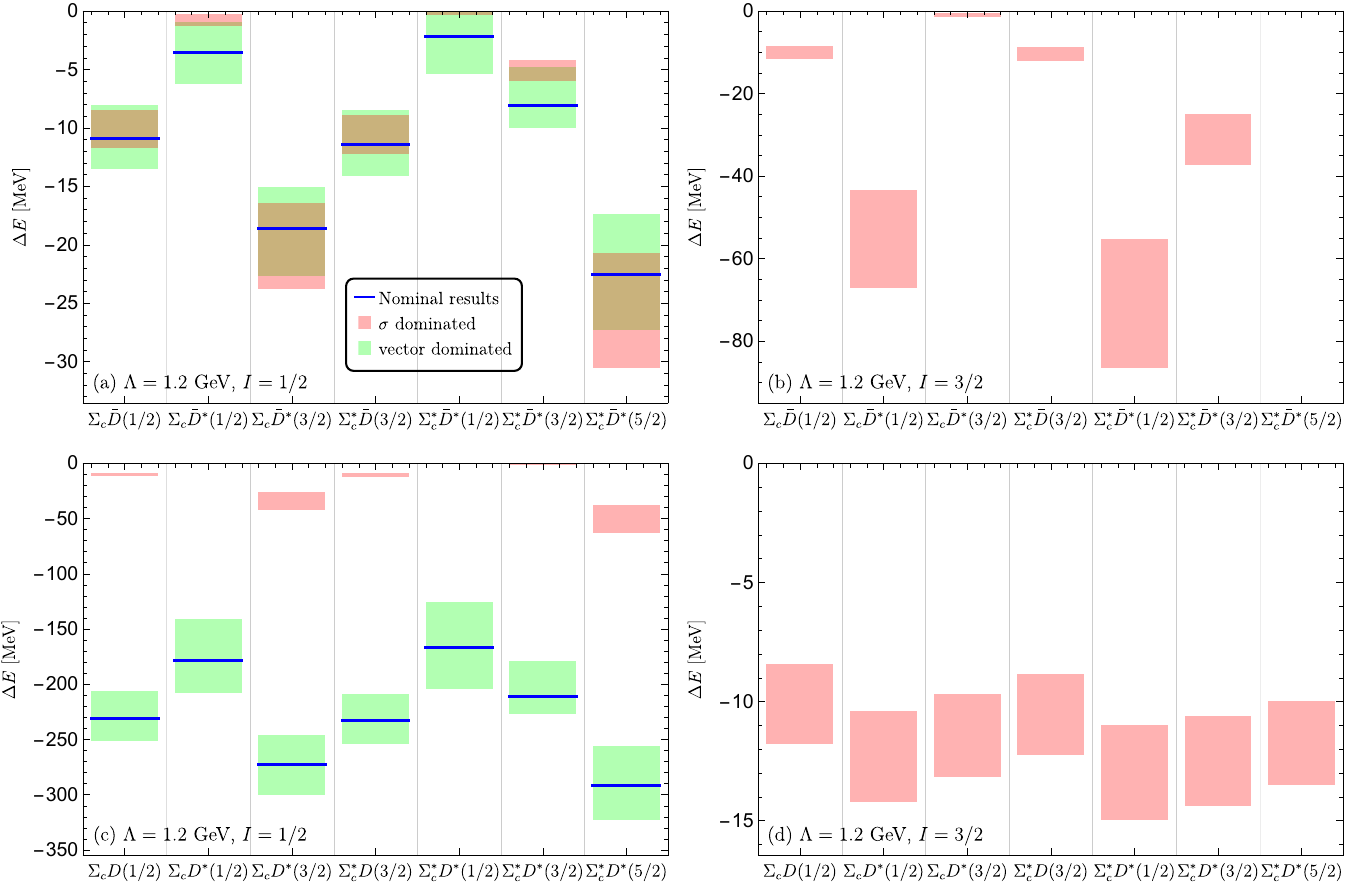}
    \caption{Mass spectrum in the OBE model with $\Lambda=1.2$ GeV with subtraction-I.}
    \label{fig:subI12}
\end{figure*}

\begin{figure*}[h]
    \centering
    \includegraphics[width=0.90\textwidth]{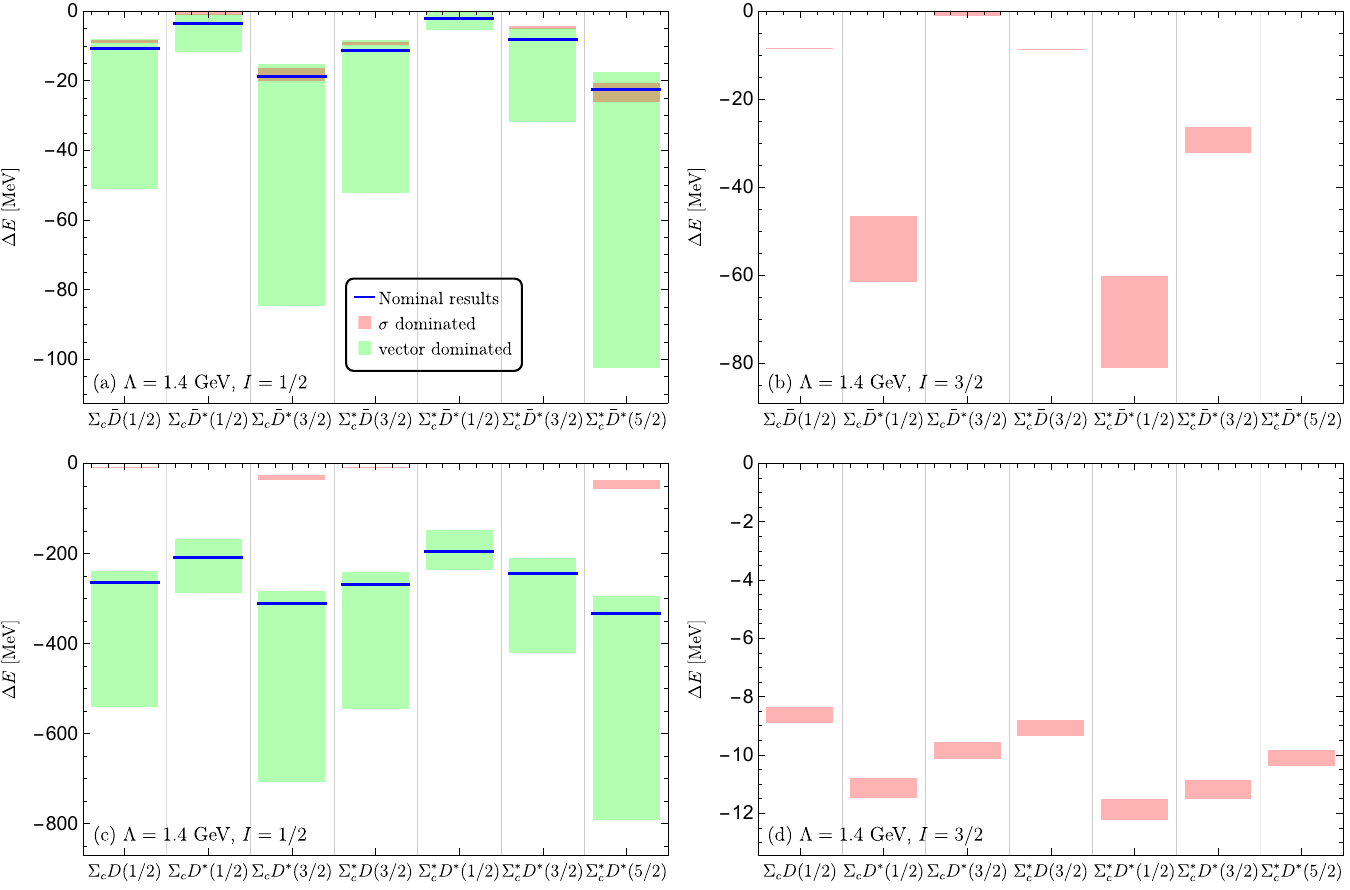}
    \caption{Mass spectrum in the OBE model with $\Lambda=1.4$ GeV with subtraction-I.}
    \label{fig:subI14}
\end{figure*}

\begin{figure*}[h]
    \centering
    \includegraphics[width=0.90\textwidth]{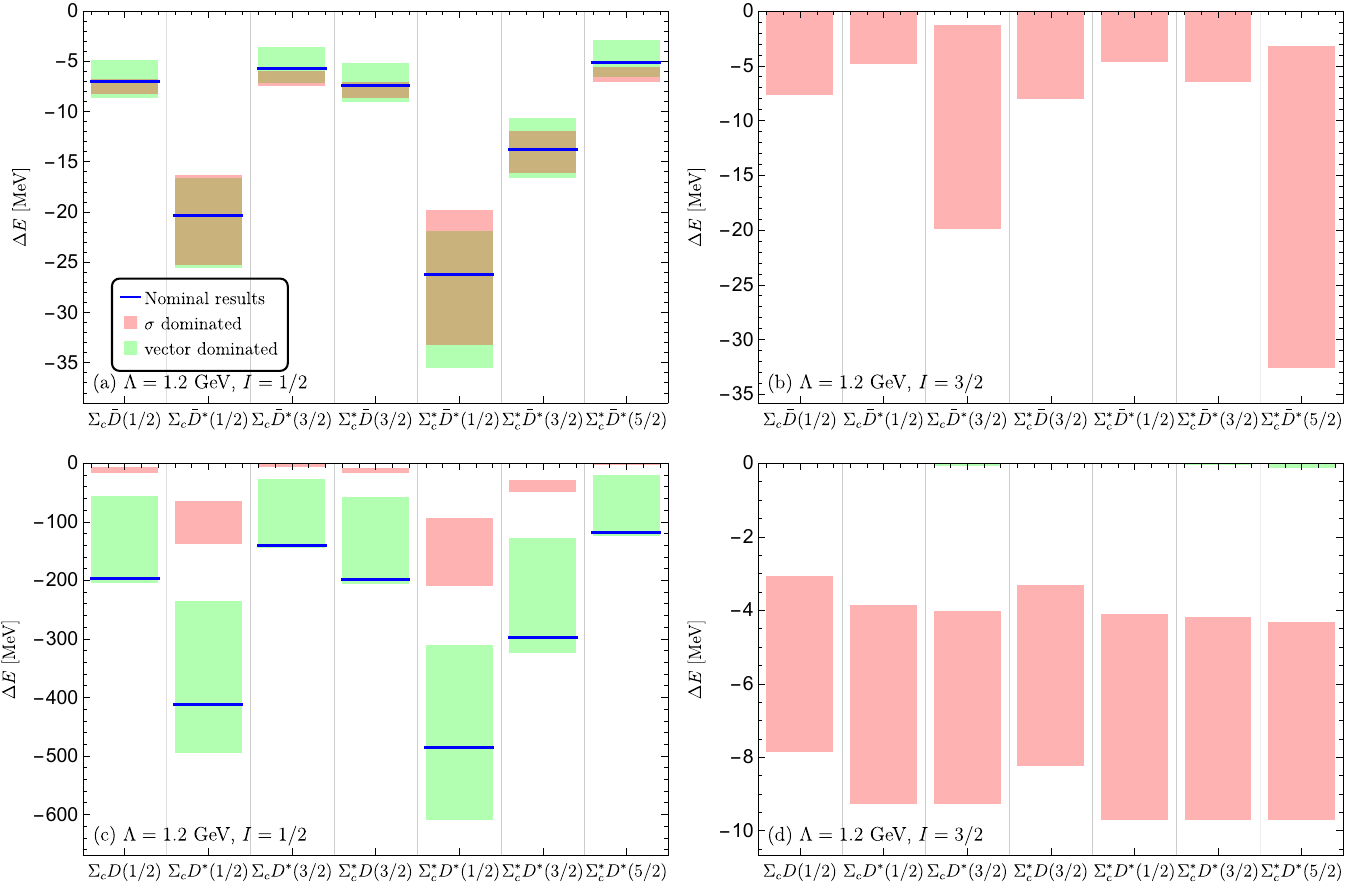}
    \caption{Mass spectrum in the OBE model with $\Lambda=1.2$ GeV with subtraction-II.}
    \label{fig:subII12}
\end{figure*}

\begin{figure*}[h]
    \centering
    \includegraphics[width=0.90\textwidth]{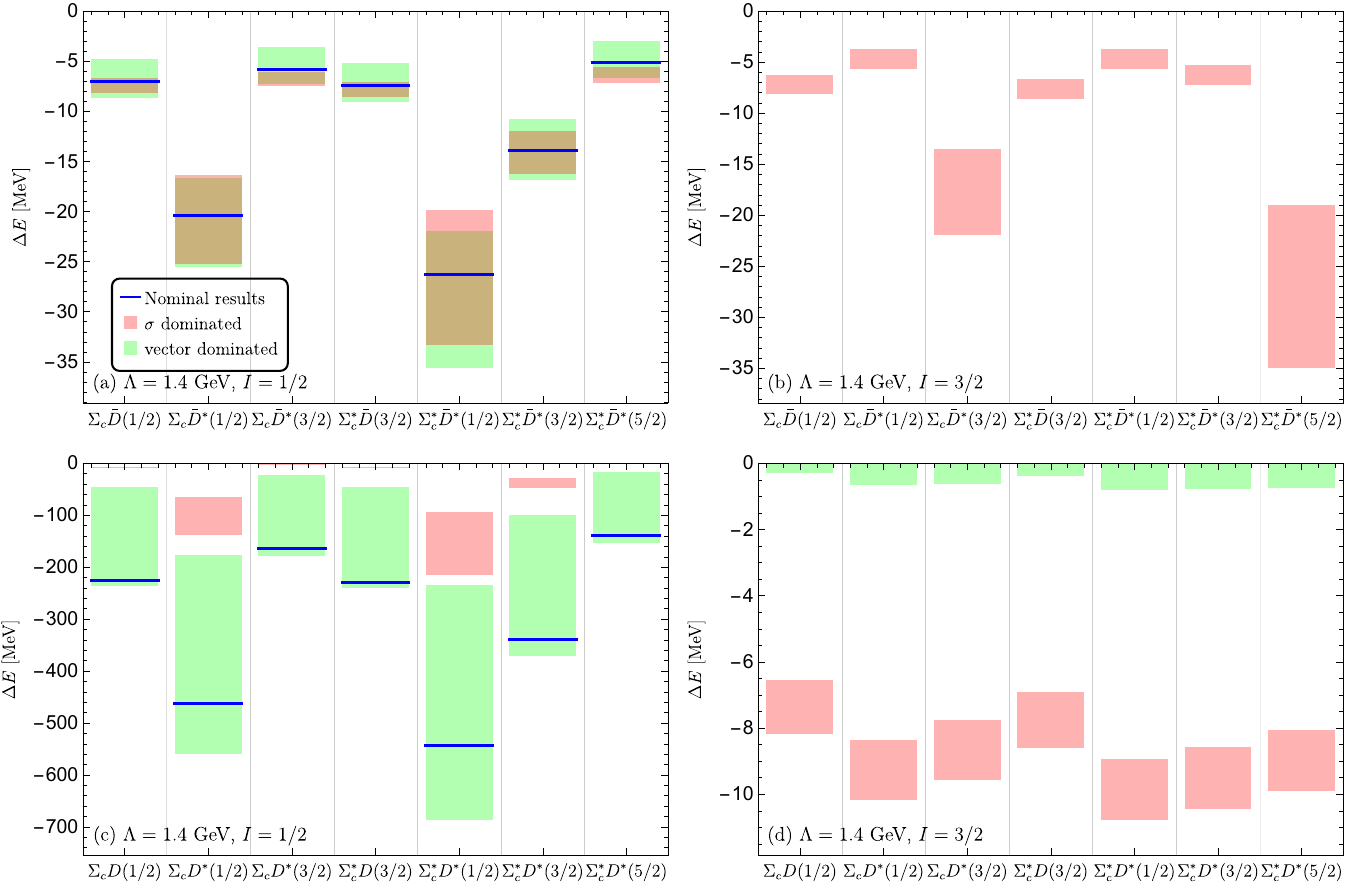}
    \caption{Mass spectrum in the OBE model with $\Lambda=1.4$ GeV with subtraction-II.}
    \label{fig:subII14}
\end{figure*}

The mass spectrum of the $\SDbar$ and $\SD$ systems in the OBE model with $\Lambda=1.2$ and $1.4$ GeV with two different subtraction schemes are given in Figs.~\ref{fig:subI12}, \ref{fig:subI14}, \ref{fig:subII12} and \ref{fig:subII14}.

\bibliography{ref}

\end{document}